\documentclass[12pt,preprint]{aastex}
\usepackage{graphics}
\usepackage{float}

\newcommand{\msun}{M_{\odot}}
\begin{document}

\title{A {\it Chandra} Study of the Rosette Star-Forming Complex.
III. The NGC 2237 Cluster and the Region's Star Formation History}

\author{Junfeng Wang,\altaffilmark{1,2} Eric
  D. Feigelson,\altaffilmark{1} Leisa K. Townsley,\altaffilmark{1}
  Patrick S. Broos,\altaffilmark{1} Carlos G. Rom\'an-Z\'u\~niga,\altaffilmark{3} Elizabeth Lada,\altaffilmark{4} and Gordon Garmire\altaffilmark{1}}

\altaffiltext{1}{Department of Astronomy \& Astrophysics, The
Pennsylvania State University, 525 Davey Lab, University Park, PA
16802}

\altaffiltext{2}{Current Address: Harvard-Smithsonian Center for Astrophysics, 60 Garden Street, Cambridge, MA 02138; {\tt juwang@cfa.harvard.edu}}

\altaffiltext{3}{Centro Astron\'{o}mico Hispano Alem\'{a}n, Camino Bajo
Hu\'{e}tor 50, Granada, Spain 18008}

\altaffiltext{4}{Department of Astronomy, University of Florida, 211 Bryant Space Science Center, Gainesville, FL 32611}

\begin{abstract}

We present high spatial resolution $Chandra$ X-ray images of the NGC
2237 young stellar cluster on the periphery of the Rosette Nebula.  We
detect 168 X-ray sources, 80\% of which have stellar counterparts in
USNO, 2MASS, and deep FLAMINGOS images.  These constitute the first
census of the cluster members with $0.2 \la M \la 2$~M$_\odot$. Star
locations in near-infrared color-magnitude diagrams indicate a cluster
age around 2~Myr with a visual extinction of $1\la A_V \la 3$ at 1.4
kpc, the distance of the Rosette Nebula's main cluster NGC 2244.  We
derive the K-band luminosity function and the X-ray luminosity
function of the cluster, which indicate a population $\sim$400--600
stars. The X-ray-selected sample shows a $K$-excess disk frequency of
13\%. The young Class II counterparts are aligned in an arc $\sim
3$~pc long suggestive of a triggered formation process induced by the
O stars in NGC~2244. The diskless Class III sources are more
dispersed. Several X-ray emitting stars are located inside the
molecular cloud and around gaseous pillars projecting from the cloud.
These stars, together with a previously unreported optical outflow
originating inside the cloud, indicate that star formation is
continuing at a low level and the cluster is still growing.

This X-ray view of young stars on the western side of the Rosette
Nebula complements our earlier studies of the central cluster NGC~2244
and the embedded clusters on the eastern side of the Nebula.  The
large scale distribution of the clusters and molecular material is
consistent with a scenario in which the rich central NGC 2244 cluster
formed first, and its expanding HII region triggered the formation of
the now-unobscured satellite clusters RMC XA and NGC 2237.  A large
swept-up shell material around the HII region is now in a second phase
of collect-and-collapse fragmentation, leading to the recent formation
of subclusters.  Other clusters deeper in the molecular cloud appear
unaffected by the Rosette Nebula expansion.

\end{abstract}


\section{Introduction}

The triggered formation of the lower mass stars in the vicinity of
massive stars is a complex process that is only now being
characterized in detail \citep[see][for a recent
  review]{Briceno07}. In their immediate neighborhood, massive stars
suppress further star formation by quickly ionizing and dispersing
surrounding molecular material \citep{Herbig62}. At greater distances,
OB stars are more constructive to star formation activity; the shocks
driven by ionization or stellar winds are crucial in triggering the
collapse of molecular cores \citep{Whitworth94,Lefloch94}.  Triggered
star formation events by massive stars have been observed at different
spatial scales, for example, small bright-rimmed clouds on the
periphery of HII regions \citep{Sugitani95, Getman07,Ogura07}, an
embedded cluster in a molecular cloud core on the edge of the
dispersed Cep~OB3b \citep{Getman06}, multiple generations of star
formation in W5 \citep{Koenig08}, a broad ridge of young stars along
the southwestern boundary of M~17 \citep{Jiang02,Broos07}, and a rich
secondary cluster on the edge of Sharpless~219 \citep{Deharveng06}.

The Rosette star forming complex has been considered an excellent
candidate for triggered star formation \citep{Cox90, PL97} following the framework developed by
\citet{Elmegreen77}.  The massive young cluster NGC 2244 \citep[$d\sim
  1.4$ kpc, $t\sim 2$ Myr;][]{Hensberge00} powers a visually
spectacular expanding HII region known as the Rosette Nebula.  The
ionized nebula is clearly interacting with the adjacent Rosette
Molecular Cloud (RMC) to the east of the Nebula which has a collection
of embedded young stellar clusters, each with a few hundred pre-main
sequence stars (Table 6 in Wang et al. 2009; see also Poulton et
al. 2008).  However, recent investigations of the ages and disk
fractions of these embedded stellar populations at near-infrared
(NIR), mid-infrared, and X-ray wavelengths do not obviously support a
sequential, triggered origin \citep{RomanZuniga08,Poulton08,Wang08a}.
Instead, Wang et al.\ suggest that the previously unnoticed, more
evolved, and less-obscured RMC~XA cluster lying between NGC~2244 and
the molecular cloud may have been triggered by the expanding Rosette
HII region $\sim 2$~Myr in the past.

To the west of NGC~2244 cluster lies the NGC~2237 cluster\footnote{The
  designations of clusters and nebulae in the Rosette region have been
  confused since the 19th century.  NGC~2237 refers both to the
  stellar cluster and its associated nebulosity on the west side of
  the Rosette nebula \citep{Sulentic73}.  The star cluster is listed
  as Ocl 512 in the catalogue of \citet{Alter70} and included in the
  study of the Monoceros star clusters by \citet{Perez91}.  NGC~2238
  and 2246 are old designations for bright emission regions around the
  Rosette Nebula; these names are rarely used today.  The well-studied
  rich star cluster exciting the nebula is NGC~2244; NGC~2239 is an
  obsolete designation for this central cluster.  The listing in the
  SIMBAD database for NGC~2237 was incorrect until recently.  Note
  that the open cluster NGC 2239 studied in \citet{Bonatto09} in fact
  refers to NGC 2237.  A thorough review of the Rosette star formation
  complex is provided by \citet{RomanZuniga08b}.}.  NGC~2237 was not
studied until the recent 2MASS and FLAMINGOS $JHK$ surveys
\citep{Li05, RomanZuniga08, Bonatto09}. Li (2005) estimates that the
cluster has a diameter of 3.2~pc with a population of 232 stars.
\citet{Bonatto09} studied NGC 2237 (called NGC 2239) with
field-star-decontaminated 2MASS photometry, and suggested that it may
be a young ($5\pm 4$~Myr) cluster located in the background of NGC
2244 ($d=3.9$~kpc). In the Digital Sky Survey optical plate
(Figure~\ref{fig:dss}), the NGC~2237 region lies on the boundary
between the HII region and the cold neutral molecular material
\citep[Core E;][]{Blitz80,RomanZuniga08b} on the western side of the
nebula. Numerous dusty elephant trunks and pillars point towards the O
stars in NGC~2244. Its structure contributes to the petal-like
morphology of the Rosette Nebula and resembles the pillared boundaries
of other HII regions such as M~16 and IC~1396 where the molecular
cloud is photoionized and ablated by the O star ultraviolet light and
winds.

However, the stellar content of NGC~2237 has not been
well-characterized and most of its members have not been individually
identified.  The difficulty is that the near-infrared surveys are
dominated by older field stars unrelated to the Rosette star forming
complex so that only the statistical enhancement in stellar surface
density, and the fraction of members with $K$-band excesses, have been
studied.  X-ray surveys are more effective at uncovering the full
pre-main sequence (PMS) population of young clusters.  Low mass young
stars emit X-rays from violent magnetic reconnection flares, orders of
magnitude more luminous than seen in older Galactic stars \citep[see
  review by][]{Feigelson07}. Hence, an X-ray study can individually
identify young stars including the diskless Class III sources missed
in the infrared-excess samples.  X-ray studies of NGC 2244
\citep[][henceforth Paper I]{Wang08a} and the embedded clusters in the
RMC \citep[][Paper II]{Wang08b} on the eastern side of the Rosette
Nebula produced a richer census of young stars than available from
infrared-excess studies.  A handful of X-ray stars in NGC~2237 have
been identified with the {\it ROSAT} satellite \citep[see][]{BC02},
but observations with the {\it Chandra X-ray Observatory} with its
high spatial resolution and sensitivity, are much more effective.

In this paper (the third in a series on the Rosette Complex), we focus
on the X-ray point sources identified in the image of the NGC~2237
region obtained with the {\em Chandra X-ray Observatory}.  We identify
over 150 individual X-ray emitting young stars and investigate their
spatial distribution and stellar properties to study the star
formation activity on the western side of the Rosette Nebula.
Proximity to dense molecular structures and correlated age-spatial
gradients in the stellar distributions are particularly valuable in
evaluating the timescale and efficiency of the triggering process that
may have occurred. We combine these results on NGC~2237 with earlier
work on the central NGC~2244 cluster and the embedded RMC clusters to
elucidate a large-scale view of the star formation history in the
Rosette complex.  A distance of $d=1.4$~kpc to NGC 2237 is adopted
throughout the paper and the uncertainty in distance
\citep[see][]{Bonatto09} will be discussed.

\section{Chandra Observations and Data Reduction \label{obs.sec}}

The NGC 2237 region was observed with the Imaging Array of the {\em
  Chandra} Advanced CCD Imaging Spectrometer (ACIS-I) which has a
field of view $\sim 17$\arcmin$\times 17$\arcmin. The 20 ks ACIS-I
observation was conducted on 9 February 2007, in the ``Timed Event,
Very Faint'' mode with 5 pixel $\times$ 5 pixel event islands (see
Table 1 in Paper I for details of pointing and roll
angle). The pointings towards NGC 2237 and NGC 2244 are outlined in
Figure~\ref{fig:dss}.

We follow the data reduction and source extraction procedures
described in Paper I.  The ACIS-I image is shown in
Figure~\ref{fig:acis}. A smoothed, exposure corrected X-ray image is
created with the CIAO tool {\it csmooth} \citep{Ebeling06} and shown
in Figure~\ref{fig:csmooth}, which gives the relative spectral
hardness of the sources. $65\%$ of the sources are dominated by soft
band (0.5-2.0 keV) X-ray emission with median photon energy $<$2~keV,
implying relatively low obscuration.  A candidate source list was
obtained using the {\it wavdetect} program \citep{Freeman02},
supplemented a few possible sources found by visual inspection.
Events are extracted and analyzed for each candidate source using our
IDL script {\it ACIS
  Extract}\footnote{\url{http://www.astro.psu.edu/xray/docs/TARA/ae\_users\_guide.html}}
\citep[version 3.128; hereafter {\it AE},][]{Broos10}.  After a
careful review of the net counts distribution for all candidate
sources, we rejected sources with greater than 1\% likelihood of
arising from a background fluctuation assuming Poisson statistics
($P_B>0.01$).  Only the brightest source in the field (\#149) has more
than 100 source counts.

The trimmed source list includes 168 valid X-ray detections, which are
further divided into a primary list of 130 highly reliable sources
(Table~\ref{tbl:primary}) and a secondary list of 38 tentative sources
with $0.001\le P_B < 0.01$ likelihood of being spurious background
fluctuations.  The resulting sources and X-ray properties are reported
in Table~\ref{tbl:primary} and~\ref{tbl:tentative} which are very
similar to the source tables in \citet{Townsley06}, \citet{Broos07},
and Paper I.  Descriptions of the table columns are given in the table
footnotes.  Six sources (\#160, 163, 164, 166, 167, and 168) are in
the ACIS region overlapping the NGC~2244 field described in Paper I.
Four of these (\#163, 166, 167, and 168) were detected in the earlier
ACIS exposure.  One bright source (\#149) was found in the {\em ROSAT}
study of \citet{BC02}.

Source event arrival times were compared to that of a uniform light
curve model in {\it AE}, and five sources (\#4, 39, 118, 149, and 152)
were found  to be significantly variable ($P_{KS} < 0.005$ in column 15 of
Tables~\ref{tbl:primary} and ~\ref{tbl:tentative}) during the short 20
ks observation. The light curve of \#149 shows a typical X-ray flare
seen in PMS stars \citep[e.g.][]{Wolk05} captured in the rise phase
near the end of the observation.

For the 65 sources with photometric significance {\em Signif}~$>2.0$
(column 12 in Tables~\ref{tbl:primary} and \ref{tbl:tentative}), the
extracted spectra were fit with a single temperature {\em apec}
thermal plasma model \citep{Smith01} subjected to an absorbing column
of interstellar material \citep[{\em tbabs} model; ]{Wilms00} with the
{\it XSPEC}\footnote{
  http://heasarc.gsfc.nasa.gov/docs/software/lheasoft/xanadu/xspec}
package \citep[version 12.2.1ap,][]{Arnaud96}.  The best-fit model was
obtained with the maximum likelihood method \citep{Cash79}. Abundances
of 0.3~$Z_{\odot}$ were assumed for the automated fitting performed by
{\it AE}.  Spectral results are presented in
Table~\ref{tbl:thermal_spectroscopy}; table notes describe the columns
and unusual cases.  Absorbing column densities range from negligible
to $\log N_H \sim 23.2$~cm$^{-2}$, equivalent to a visual absorption
of $A_V \sim$~100~mag \citep{Vuong03}. Temperatures range from $kT
\sim 0.1$~keV to $kT >15$ keV; this truncated value reflects the
hardest spectra detectable by the ACIS detector.  The range of total
band ($0.5-8$~keV) absorption-corrected luminosities derived from
spectral modeling is $30.0 \lesssim \log L_{t,c} \lesssim 32.1$
ergs~s$^{-1}$.

To estimate the X-ray sensitivity of the ACIS exposure, we use the
Portable, Interactive Multi-Mission Simulator (PIMMS)
software\footnote{\url{http://heasarc.gsfc.nasa.gov/docs/software/tools/pimms.html}}
for a limiting on-axis source with 3 counts assuming a 2~keV
plasma temperature and an absorbing column $\log N_H \sim
21.4$~cm$^{-2}$ (corresponding to $A_V=1.5$~mag visual
extinction).  The resulting limiting total band luminosity as
observed by the telescope is $\log L_t \sim 29.4$~ergs~s$^{-1}$ and
limiting intrinsic luminosity $\log L_{t,c} \sim
29.6$~ergs~s$^{-1}$.  A conservative estimate for limiting
luminosity of the entire NGC 2237 observation, including
sensitivity degradation due to off-axis optics and higher
absorption, is $\log L_t \sim 30.0$ erg s$^{-1}$.  Using the
empirical correlation between X-ray luminosity and mass in PMS
stars from well-studied samples in the Orion and Taurus clouds
\citep{Preibisch05,
  Telleschi07}, we infer that the observation is complete to all
cluster members down to $M \sim 1.0$~M$_{\odot}$ and includes an
incomplete fraction of cluster members down to $M \sim 0.2$~M$_\odot$.

\section{Stellar Counterparts to {\em Chandra} Sources}\label{sec:counterpart_sec}

\subsection{Correlation of X-ray and Optical/Infrared Catalogs}

We searched optical and infrared catalogs from the literature and
recent observations for stellar counterparts to these 168 X-ray
sources. In addition to the available catalogs listed in Paper II
(USNO, 2MASS, and FLAMINGOS), other catalogs are: $UBV$ photometry of
NGC~2244 (Ogura \& Ishida 1981, OI81; Massey et al. 1995,
MJD95), $BVIR$H$_{\alpha}$ photometry of NGC~2244 (Bergh\"{o}fer \&
Christian 2002, BC02).  {\it Spitzer Space Telescope} mid-infrared
observations are not yet available for this field.  The reference
frame offsets between the ACIS field (which is astrometrically aligned
to the Hipparcos frame using 2MASS sources in the data reduction) and
the catalogs are $0.3^{\prime\prime}$ to OI81, $0.4^{\prime\prime}$ to
MJD95, $0.3^{\prime\prime}$ to BC02, $0.2^{\prime\prime}$ to USNO, and
$0.2^{\prime\prime}$ to FLAMINGOS.  These offsets were applied before
matching sources.

We associate ACIS X-ray sources with optical and near IR (ONIR)
sources using positional coincidence based on the positional
uncertainties of the X-ray sources and stellar catalogs
\citep{Broos07}. Likely associations for 134 of the 168 ACIS sources
are given in Table \ref{tbl:counterparts} along with ONIR photometric
data (see Table \ref{tbl:counterparts} footnote for photometric
errors).  Typical offsets between the X-ray and near-infrared sources
are 0.2\arcsec\/ for on-axis sources and 0.7\arcsec\/ for off-axis
sources.  Although USNO photometry is reported when optical
counterparts are found, readers are cautioned that magnitudes are
derived from photographic plates with $\sim 0.3$ magnitude photometric
accuracy \citep{Monet03}. $JHK$ magnitudes from 2MASS are given when
the source falls outside the FLAMINGOS field of view. The SIMBAD and
VizieR online catalogs are searched for complementary information with
results given in the table footnotes.

The $Chandra$ observation is able to identify a majority of the
estimated $\sim 230$ members of the NGC~2237 cluster \citep{Li05}.
This excellent result arises from a combination of three effects: the
short ACIS exposure is sufficient to detect a large fraction of the
stellar initial mass function (IMF), even with significant obscuration
(\S \ref{obs.sec}); the catalog from FLAMINGOS deep $JHK$ imaging is
sensitive to most of the cluster population; and the X-ray image
suffers only minor contamination from non-cluster X-ray sources
(\S~\ref{sec:no-id}).

\subsection{Near-infrared and Disk Properties}\label{nir_disk}

The $J-H$ vs.\ $H-K$ color-color diagram for 119 {\em Chandra} stars
is shown in Figure~\ref{fig:ccd}. This sample is restricted to stars
with high-quality $JHK$ photometry (i.e., errors in both $J-H$ and
$H-K$ colors $<0.1$ mag). Most {\em Chandra} sources occupy the color
space associated with diskless young stars, Class III or weak-lined T
Tauri stars (WTTS), that are reddened by interstellar extinction.  A
concentration of cluster members subjected to $A_V\sim 2$~mag is
apparent, approximately centered at $J-H=0.80$ and $H-K=0.35$. This is
also the typical extinction seen towards NGC 2244 \citep[e.g.,
][]{OI81,BC02}. Seven stars with $K$-band excesses lie to the right of
this reddening band\footnote{These sources are \#44, 61, 80, 97, 111,
  117, and 121. For the two sources (\#111 and \#121) that are close
  to the reddening band, the errors in both $J-H$ and $H-K$ colors are
  small, $\sim 0.015$ mag, therefore their $K$-excesses are small but
  statistically significant. See also Paper I for a brief discussion
  on systematics in the $K$-excess.  These stars are also present in
  \citet{RomanZuniga08}.}, defined as stars that have colors
$(J-H)>1.7(H-K)+2\sigma(H-K)$. These are likely PMS stars with
circumstellar accretion disks, Class II objects or classical T Tauri
stars (CTTS).  One outlier among the Class III sources, \#86, appears
highly obscured with $A_V \sim 20$ mag. Among the faintest X-ray
sources with only 4 net counts in the ACIS image, this source is
located in an optically dark region; this source is discussed in
\S~4.2.  We did not detect the red sources in \citet{RomanZuniga08}
with no detection in $J$ and $H-K> 1.5$ mag, the majority of which
were located within the molecular cloud (cf. Figure 27 in
Rom\'an-Z\'u\~niga et al.).  However, four ACIS sources (\#45, 70, 73,
85) have only $K$-band counterparts with no $J$ and $H$ detections, and
hard median photon energy ($\sim$2.5 keV).  They are probably deeply
embedded young stars, similar to those red sources.

Figure~\ref{fig:cmd} shows the NIR $J$ vs.\ $J-H$ color-magnitude
diagram (CMD) for the same stars presented in
Figure~\ref{fig:ccd}. The 1 Myr and 2 Myr PMS isochrones from the
evolutionary tracks of \citet{Siess00} are shown for reference. If the
NGC 2237 satellite cluster was triggered by the central NGC 2244
cluster, it must be younger than NGC 2244 at $\sim 2$ Myr
\citep{Hensberge00}. While it is not possible to directly measure the
age of NGC 2237 without obtaining spectra of the cluster members, the distribution
of stars in NGC 2237 in the NIR color-magnitude diagram is nearly
identical to that seen in the main NGC 2244 cluster (cf.\ Figure 6 of
Paper I).  Approximate stellar masses can be measured from the
iso-mass vectors (blue dashed lines) on the color-magnitude diagram.
They range from early-B stars to early-M stars.

From Figures~\ref{fig:ccd} and~\ref{fig:cmd}, we find that the
fraction of K-band excess sources is $13\%$ among stars with $M \ga
1$~M$_\odot$, where we estimate masses from the NIR color-magnitude
diagram (recall this is also the completeness limit of the
X-ray-selected stars).  This closely matches the $K$-excess disk
fraction of $15\pm 3\%$ reported in the FLAMINGOS NIR study of NGC
2237 \citep{RomanZuniga08}.  However, when the full X-ray sample
including many sub-solar mass stars is considered, the $K$-band excess
fraction of the X-ray sample is 6\%.  Since disks around intermediate-mass stars evolve
faster than disks around low-mass stars (e.g.,
Hern{\'a}ndez et al. 2007), most likely this is a selection effect
against the X-ray emission from CTTS which are known to be
systematically lower than the X-rays from WTTS by a factor of 2 (e.g.,
Preibisch et al. 2005).

\subsection{Non-cluster Contaminants and Embedded Sources \label{sec:no-id}}

We deduce from previous studies (e.g., Wang et al. 2007) that the 34
ACIS sources without any matched counterparts are a mixture of newly
discovered embedded cluster members, distant background stars, and
extragalactic sources. The level of contamination by extragalactic
X-ray sources and Galactic disk stars is evaluated with simulations as
described in Paper I.  The extragalactic population is simulated by
placing artificial sources randomly across the detector with incident
fluxes drawn from the X-ray background $\log N$--$\log S$ distribution
\citep{Moretti03}.  The Galactic stellar population expected in the
direction of our ACIS field is simulated with the stellar population
synthesis model of \citet{Robin03}, adopting X-ray luminosity
distributions of nearby stars measured from $ROSAT$ surveys
\citep[e.g., ][]{Hunsch99}.  These simulations predict that $\sim 12$
sources will be extragalactic and $\sim$10 could be old field
stars. These contaminants constitute $\sim 13\%$ of the 168 ACIS
sources.  Figure~\ref{fig:hard} compares the spatial distribution of
X-ray sources without NIR counterparts and sources that have hard
median photon energies ($E_{med}>2.0$ keV). The 10 unidentified X-ray
sources that are located in the unobscured HII region cavity should
have counterparts if they are cluster members.  These sources, plus a
few in the obscured region, are most likely the background
extragalactic sources and background field stars.  The other 24 ACIS
sources lie in the molecular pillars and the optically darker regions
with dense molecular material.  They are likely new embedded low-mass
PMS stars that are fainter than the FLAMINGOS completeness limit. Such
X-ray-discovered stars are commonly found in the molecular clouds
surrounding other clusters \citep[][Paper I]{Getman05a, Broos07}.

\subsection{Properties of Selected Stars}

A few stars have bright $J$ magnitudes indicating they are
intermediate-mass stars (B and A types) reddened from the ZAMS with $0
< A_V < 2$~mag. Their locations in the color-color diagram
(Figure~\ref{fig:ccd}) are also consistent with early spectral types,
and their bright optical counterparts are identified by OI81, MJD95 or
other studies.  Two of these sources (\#44 and \#61) have masses $M>2
\msun$ estimated from the color-magnitude diagram and strong $K$-band
excesses; these are good candidates for being accreting Herbig Ae/Be
(HAeBe) stars.  Others do not show NIR excess.  Wang et al. (2007)
also find that in massive OB cluster NGC 6357, over 90\% of the X-ray
selected intermediate-mass stars are not HAeBe stars.  Perhaps this is
related to the fast disk evolution in the HAeBe stars
\citep[e.g.,][]{Waters98}, or an X-ray deficiency in detecting the
intermediate stars with NIR-excess disks \citep[e.g.,][]{Hamaguchi05}.

ACIS source \#53 is a B1 star \citep{Philip80}, the earliest spectral
type known in the NGC 2237 region. It only has 16 net counts and the
spectral fitting gives $\log L_{t,c} \sim 30.2$~erg~s$^{-1}$. A B2
star identified by \citet{Philip80} is source \#54 in our X-ray study.
With over 100 counts, its X-ray spectrum can be reasonably
well-determined (Table~\ref{tbl:thermal_spectroscopy} and
Figure~\ref{fig:spec}a): it is dominated by a plasma with moderate
temperature ($kT=1.2$~keV) with low absorption ($\log N_H \sim
20.0$~cm$^{-2}$) and X-ray luminosity $\log L_{t,c} \sim
31.0$~erg~s$^{-1}$ (0.5-8 keV).  These X-ray properties are similar to
those of other early-B stars; for example, in the Orion Nebula Cluster
\citep[ONC;]{Stelzer05} and NGC~2244 Rosette Cluster (Paper I).

Source \#149 is the X-ray source with most net counts in this study
with a spectrum well-fit by a single plasma of $kT=4.1$ keV and
absorption column of $\log N_H \sim 21.3$ cm$^{-2}$
(Figure~\ref{fig:spec}b). The $kT$ value is higher than that of a
typical PMS star \citep[e.g.,][]{Gudel07}, caused by the X-ray flare
seen near the end of the observation. It is known from previous
observations that spectral hardening is commonly seen during X-ray
flaring in PMS stars \citep[e.g.,][]{Getman08a}. The absorption
corrected total band (0.5--8 keV) X-ray luminosity $\log L_{t,c} \sim
31.3$~erg~s$^{-1}$.  \citet{BC02} reported a similar level of $\log
L_x=31.01$ erg s$^{-1}$ (0.5--2 keV) in their {\em ROSAT}/PSPC
observation.  This source represents one of the typical flaring PMS
stars seen in the ONC \citep{Wolk05,Getman08a} and many other young
star clusters \citep{Wolk08}.

Lastly, ACIS source \#20 is very interesting in its X-ray properties,
showing a high X-ray luminosity ($\log L_{t,c}\sim 31.3$ erg
s$^{-1}$), hard spectrum ($kT\sim 3.4$ keV), and high absorption
($N_H\sim 10^{23}$ cm$^{-2}$). It is located in the optically dark
pillar region at the peripheries of the HII nebula.  No counterpart is
found in the 2MASS image and the FLAMINGOS image, and no information
on its mid-IR or mm band properties is currently available.  It may
represent a similar detection of Class 0/I protostar as the VLA/BIMA
source IRAS 21391+5802 in IC 1396N (Getman et al. 2007) or a chance
superposition of a background AGN.

\section{Cluster Properties}\label{sec:ch6_spatial.sec}

\subsection{Cluster Center}

The central position of the NGC 2237 cluster has not been
well-determined from NIR observations.  \citet{Li05} obtains
($\alpha$,$\delta$) = (6$^h$30$^m$55.8$^s$,
+04$^\circ$58\arcmin30\arcsec) from a weak 2MASS surface density
enhancement, while \citet{RomanZuniga08} obtains (6$^h$30$^m$22$^s$,
+04$^\circ$56\arcmin) from the $K$-band excess sources in the deeper
FLAMINGOS data \citep[see Table 1 in ][]{RomanZuniga08}.  The 2MASS
position is $\sim$8\arcmin\/ east of the FLAMINGOS position.

The X-ray position for the cluster core and the spatial distributions
of stellar populations are best examined from a smoothed map of the
X-ray selected cluster members.  Following our Papers I and II, we
derived a smoothed map of the surface density of the X-ray-selected
cluster members using a 1.5\arcmin\/-radius top-hat smoothing kernel
(Figure~\ref{fig:ssd}).  To reduce the effect of sensitivity
variations across the field, we included only 93 sources that have 6
or more net counts.  Adopting different kernel widths does not have
a significant effect on the resulting features discussed here.  The
stellar density in the western portion of the field with molecular
cloud material is likely underestimated compared to the eastern
portion for two reasons: 24 unidentified X-ray sources are probably
embedded stars (\S~\ref{sec:no-id}) and the sensitivity to embedded
stars is reduced due to soft X-ray absorption.

A clear peak stellar concentration appears at (6$^h$31$^m$00$^s$,
+04$^\circ$58\arcmin).  This is the most accurate available as the
X-ray contour map has a much higher dynamic range due to its smaller
contamination by extraneous sources.  This position derived from the
$Chandra$ selected stars is consistent with the low-resolution {\em
  ROSAT} X-ray image shown by \citet{BC02}, and agrees reasonably well
with 2MASS result, lying 1\arcmin\/ towards the main NGC~2244 cluster.
The discrepancy in the infrared cluster centroids is due to different
fields of view, definition of cluster center, and possibly age
effects: the FLAMINGOS field extends $\sim 10$\arcmin\/ westward of
the ACIS field, and \citet{RomanZuniga08} choose to define the cluster
to encompass Class~II systems found in this extended region, while the
X-ray map and the 2MASS $K$ surface density are weighted toward Class
III sources.  Our X-ray star distribution shows a secondary peak
8\arcmin\/ northwest of the primary peak (see
\S~\ref{sec:concentration}), which is close to the NGC 2237 core
position defined in \citet{RomanZuniga08}.

\subsection{Spatial Distribution of the Young Stars}\label{sec:concentration}

The spatial distribution of cluster members has not been
well-established from NIR studies due to the high contamination
level of non-cluster stars.  The diameter of the NGC~2237 cluster
has been estimated to be $\sim 3.2$~pc (8\arcmin) from the 2MASS
enhancement \citep{Li05} and 3.8~pc from the $K$-band excess stars
\citep{RomanZuniga08}.  Our X-ray-selected NGC 2237 stars, which
represent the bulk of cluster members, indicate that the structure
is more complex than a spherical aggregate with a single size.

From Figure~\ref{fig:ssd}, we see that the dimension of the main
stellar concentration around the cluster core at
(6$^h$31$^m$00$^s$, +04$^\circ$58\arcmin) is $\sim 6\arcmin \times
3\arcmin$ ($\sim 2.5 \times 1.3$~pc) in diameter, asymmetrically
elongated towards the west from the dense core.  A sparse
secondary stellar concentration may also be present: a
northwestern group of $\sim 15$ stars in a $1.5 \times 1$~pc
region around (6$^h$30$^m$39$^s$, +05$^\circ$00\arcmin).

We next investigate whether any spatial segregation is seen in stars
at different stages of PMS evolution.  Figure~\ref{fig:spatial} shows
the distribution of X-ray-selected Class II and Class III sources
classified using the near-infrared color-color diagram
(Figure~\ref{fig:ccd}). The 34 sources without matched counterparts,
many of which are probably embedded stars (\S~\ref{sec:no-id}), are
also shown. The seven Class II sources appear to be aligned along an
arc parallel to the edge of the dusty cloud edge and perpendicular to
the direction towards the early O stars in NGC 2244. Most of the Class
III sources are associated with the groupings mentioned above; the
southern and eastern edges of the field seem to lack X-ray sources.
The CO emission contours suggest that there is some molecular material
to the west.  The sensitivity of our observation is too shallow to
probe the star formation activity here because of heavy absorption.

Figure~\ref{fig:mass} shows the spatial distribution of stars in
different mass ranges. We consider stars with identified B and A
spectral types (see Table~\ref{tbl:counterparts} footnote) and stars
with NIR-estimated masses $\ga 2\msun$ as the high mass sample, while
stars in the mass range $M< 2\msun$ are the low mass sample. The high mass
stars are spatially concentrated at the cluster center.  Note that
sources \#44, 53, and 61 lie along the same arc as the $K$-band excess
Class II sources. In contrast, the majority of the low mass stars are
not strongly clustered, but are dispersed along the peripheries of the
HII region and inside the dark cloud. Again, the deficit of sources in
the southern and western edges of the field is worth noting.

A close-up H$\alpha$ image \citep{Li04} of a small region at the
interface between the HII region and the cloud is shown in
Figure~\ref{fig:halpha}. A number of X-ray stars are found at or near
the tip of dusty pillars (ACIS \#77, 86, 91, 96, 100, and 101) in the
two regions outlined by circles.  The {\em Chandra} images of the
Eagle Nebula (M~16) and NGC 6357 are similar with a handful of
hard-spectrum X-ray stars embedded in dusty pillars on the edges of
HII regions ionized by rich clusters (Linsky et al.  2007, Wang et
al. 2007).  It is plausible that these stars become exposed because
the OB stars of NGC 2244 (to the east) have photoevaporated the
surrounding cloud material.  The most obscured source, \#86, is
embedded in or behind the southern pillar shown.  Only 4 counts are
seen from this source, most above 2~keV. \citet{Li05} also noted a
highly obscured 2MASS source with extinction $A_V\sim 17$ mag located
behind the same pillar object, however no matching X-ray source was
detected in our observation, likely due to the limited sensitivity.

Another interesting feature in this region is a bright visually
identified outflow outlined by the white box in
Figure~\ref{fig:halpha}.  Located at (6$^h$30$^m$48.7$^s$,
+05$^\circ$01\arcmin44\arcsec), it has not been reported previously in
the literature, to our knowledge.  This may be a Herbig-Haro (HH)
object.  There is a visible star 30\arcsec/\ to the northwest aligned
with the outflow which might be its host star. No X-rays are detected
from the candidate ionizing star or from the jet.

We thus find that the spatial distribution of the {\em Chandra} stars
(Figure~\ref{fig:ssd}) and their NIR counterparts
(Figure~\ref{fig:spatial}) show the following features: a compact
cluster previously recognized as NGC~2237 which is $\sim 2.5\times
1.3$ pc in size, elongated along an east-west axis parallel to the
direction of the NGC 2244 early O stars.  Its shape might reflect an
origin in a cometary cloud that was triggered to form stars and
ablated by the O star ultraviolet photons and winds.  A broader
distribution of stars with a possible sparse subcluster is seen over a
$5\times 5$~pc region northwest of the main cluster, partly in the
evacuated Rosette HII region and partly embedded in the optically dark
pillars and cloud.  The young star density is distinctly reduced
$1-3$~pc south and west of the cluster; the western deficit can be
attributed to absorption from the residual molecular cloud material.
This NGC~2244 satellite cluster contains about ten intermediate-mass
stars, consistent with its estimated population from the 2MASS study.
There are hints of mass segregation (the intermediate mass stars lie
primarily in the HII region) and age segregation (the $K$-band excess
stars like primarily along the cloud edge), but a clear age sequence
is not seen.  We note that richer substructures in the stellar
distributions of massive young clusters have been seen in our X-ray
studies of NGC~6357 \citep{Wang07}, M~17 \citep{Broos07}, NGC~2244
(Paper I), and NGC 6334 \citep{Feigelson09}.

\subsection{K-band and X-ray Luminosity Functions}\label{klf_xlf}

Following \citet{Li05} and \citet{Wang08a} we create the apparent
$K_s$-band luminosity function (KLF) for NGC 2237 with the spatially
complete 2MASS data, subtracting the KLF of a control field
representative of the background distribution of NGC 2237 and reddened
with $A_V=2$ mag.  Our KLF, shown in Figure~\ref{klf}, is consistent
with the one derived for NGC 2237 in \citet{Bonatto09}, which turns
over beyond the completeness limit of 2MASS \citep[$K_s=14.3$
  mag;][]{Skrutskie06}. It is well fit with a power-law distribution
in the range $ 12 < K_s < 14.3$ mag, giving a slope of $\sim$0.6.
Within uncertainties, this is marginally consistent with model
predictions of $\sim$0.4 for clusters of similar ages \citep{Lada95}.
A significant excess of stars is seen as a ``bump'' at $K_s=11.5$ in
the KLF \citep[see also Figure 12 in][]{Bonatto09}, indicating the
presence of more intermediate-mass stars than expected from a simple
extension of the KLF slope.  This perhaps reflects the contribution
from a secondary cluster in the field or foreground/background stars.
For comparison, the KLF for the young, embedded cluster RMC XC (Paper
II) is also derived and shown in Figure~\ref{klf}.  No excess of
$K$-bright stars is seen in the KLF and the slope is 0.39, in
agreement with the KLF slopes in \citet{Lada95}.

We further examined the 13 $K$-bright stars to identify their nature.
The contamination of bright stars from NGC 2244, the central cluster
in the Rosette Nebula, is too small to account for the $K$-bright
stars found.  Radial profiles of NGC 2244 (Figure 2 in Li 2005; see
also Paper I) show that the stellar surface density of NGC 2244 is
approximately 0.2 star per square arcmin at a distance $r\sim
15\arcmin$ from the cluster center and has already decreased to zero
at $r\sim 19\arcmin$.  At most two NGC 2244 stars may be present in
the NGC 2237 region, which would not necessarily be $K$-bright stars
even if they are detected.

Three of these $K$-bright stars are not detected as X-ray sources, so
they are most likely nearby foreground stars along the line of sight
towards NGC 2237.  For the remaining 10 X-ray-detected K-band bright
sources\footnote{These sources are ACIS\# 44, 53, 54, 59, 71, 83, 128,
  129, 138, and 141.}, their median photon energy ranges between 1 and
1.5 keV.  None shows very soft emission (median energy $\ll 1$ keV)
that is suggestive of foreground stars with low absorption.  Using $B$
and $V$ magnitudes from the NOMAD catalog \citep{Zacharias05}, we
derived the optical CMD for these stars, shown in Figure~\ref{10src}.
Adopting the tabulated spectral types and photometric properties of
stars in \citet{SK82}, the zero-age main sequence (ZAMS) isochrone is
shown, assuming a minimum reddening of $E(B-V)$=0.5 and a distance of
$d=1.4$ kpc for NGC 2237 members.  Four stars (\#54, 59, 83, 138)
appear to be likely foreground $FGK$ stars; the ZAMS location at $d=1$
kpc with no reddening is shown for reference.  The rest are consistent
with being intermediate-mass stars in NGC 2237.

Similarly we derived the observed hard band XLF \citep[see][]{Wang08a}
for NGC 2237, which is more robust against the inhomogeneous
obscuration.  Luminosities were taken from the spectral fits for the
brighter X-ray sources, and estimated with a simple absorbed thermal
plasma model for the faintest sources with few counts.  We adopt an
$N_H=4\times 10^{21}$ cm$^{-2}$ (derived from $N_H/A_V=1.6\times
10^{21}$ cm$^{-2}$ mag$^{-1}$; Vuong et al. 2003, and a typical
measured $A_V\sim 2.5$ mag from Figure~\ref{fig:ccd}), and a $kT=2$
keV based on the median photon energy of the faint sources
\citep{Feigelson05}.  The XLF with a farther distance $d=3.9$ kpc was
also generated similarly, except that a higher $N_H=5.5\times 10^{21}$
cm$^{-2}$ is adopted (Bonatto \& Bica 2009).

Figure~\ref{xlf} shows the resulting XLF. The uncertainties for the
hard band luminosity $L_h$ derived from spectral fits range between
0.1 and 0.3 dex.  To include both the $\sqrt{N}$ counting error and
the $Lx$-dependent measurement error, following Getman et al. (2006),
XLF error bars shown in Figure~\ref{xlf} cover the range containing
68\% ($1\sigma$ equivalent) of the Monte Carlo simulated XLF
distributions, where individual X-ray luminosities are randomly drawn
from Gaussian distributions with mean equal to the measured $\log Lh$
and variances $\sim$0.3 dex (observed $\Delta\log L_h$).

For comparison, the XLF for NGC 2244 and the XLF for ONC from the
Chandra Orion Ultradeep Project \citep[COUP][]{Getman05b} were
included.  Considering $\log L_h\ga 29.6$ erg s$^{-1}$, where the
detections are mostly complete, the slope of the NGC 2237 XLF matches
the ONC XLF well, but a scaling factor $\sim$0.2--0.3 is required to
fit the ONC X-ray-detected population.  Given that the ONC contains
$\sim$2000 members with mass above the stellar limit
\citep[e.g.,][]{Hillenbrand97}, this implies a smaller cluster
population $\sim$400--600 for NGC 2237.  The X-ray-inferred population
agrees well with the NIR population of $\sim$230 stars above 0.8
M$_{\odot}$, which is $\sim$40\% of the total population assuming a
standard IMF.  In addition, comparison of the NGC 2237 KLF to the RMC
XC cluster KLF (Figure~\ref{klf}) also supports that NGC 2237 is less
rich than the RMC XC cluster, which has an estimated population of
$\sim$800 stars (Paper II).  We conclude that NGC 2237 contains
$\sim$400--600 members.

\subsection{Discussion on the Distance to NGC 2237}

The adopted distance to NGC 2237 in this work is $d\sim 1.4$~kpc,
consistent with the generally accepted distance to the Rosette region
\citep[e.g.,][]{OI81,RomanZuniga08b}.  A spectroscopic survey is
needed to securely establish the distance to NGC 2237.  In
\citet{Bonatto09}, the decontaminated near-infrared CMD is not well
constrained and allows alternative age and distance solutions.  For
assumed ages of 1~Myr and 5~Myr, they derived best-fit distances of
$d=3.9\pm 0.4$ kpc and $d=1.7\pm 0.3$ kpc, respectively.
\citet{Bonatto09} favor the further distance with an age $5\pm 4$ Myr
since it better accounts for the decontaminated CMD sequence.  NIR
photometry of the B5 star GSC00154-02384 agrees with the larger
distance ($d=4.3$ kpc) while photometry of the A2 star GSC00154-01659
agrees with the closer distance ($d=1.4$ kpc).  We note that the
spectral types were estimated from optical $uvby$ photometry
\citep{Philip80} and may be uncertain.  The A2 and B5 star are also
detected in our Chandra observation as ACIS \#128 and \#138,
respectively.  Both have $\sim 10$ net counts, and the full-band
absorption-corrected X-ray luminosities $\log L_{t,c}\sim 30-31$ (for
$d=1.4$ kpc) are acceptable for intermediate-mass stars, as the X-ray
emission often comes from their lower-mass companions
\citep[e.g.,][]{Stelzer05}.

\citet{Getman08b} proposed an X-ray method to determine distance to
young star clusters, based on two interlocking features: X-ray
luminosities of PMS stars depend on stellar mass, as seen in a wide
variety of PMS populations independent of the detailed age
distribution and star formation mode (Preibisch et al.\ 2005;
Telleschi et al.\ 2007); and X-ray luminosities and masses scale
differently with the assumed distance.  We attempted this method to
obtain an independent estimate of the distance to NGC 2237 using a
reliable sample of bright and less obscured ($N_H<10^{21}$ cm$^{-2}$)
X-ray sources with $L_{t,c}$ derived in
Table~\ref{tbl:thermal_spectroscopy} and mass $M$ derived from NIR
photometry and Siess et al. (2000) tracks.  However, comparison with
the well calibrated relation $L_{t,c}=1.99\log M+30.43$ erg s$^{-1}$
for the Cepheus OB3b and Orion stars (Getman et al. 2006) does not
provide a decisive distance for the NGC 2237 cluster: a 2 Myr old
cluster at $d=1.4$ kpc and a 5 Myr old cluster at $d=3.9$ kpc both
follow the relation with comparable scatter.

The XLFs discussed in \S~\ref{klf_xlf} are more effective and support
a nearer distance.  Using $d=3.9$ kpc, the NGC 2237 XLF implies that
it is a rich massive cluster $\sim 1.5-2$ times the population of the
ONC and NGC 2244.  Dozens of OB stars should be present for a standard
IMF, and these are absent in the optical, NIR, and X-ray images.  In
addition, the X-ray-measured absorption columns are moderate, a few
times $10^{21}$ cm$^{-2}$, consistent with $A_V\sim 1-3$ mag seen
towards the central Rosette NGC 2244 cluster.  For a distance $d=3.9$
kpc along the Galactic plane, a significantly higher $A_V$ is
expected.

Therefore, we consider our choice of the near distance justified.
Note that the larger distance to NGC 2237 is not ruled out by the rich
population estimated from the XLF comparison; an anomalous IMF could
explain the lack of OB stars.  Further optical spectroscopic surveys of the
NGC 2237 cluster will be invaluable to confirm whether it is a
background cluster behind the Rosette region at a distance of $d=3.9$
kpc.

\section{The Global View of Star Formation Across the Complex}

\subsection{The Relation of NGC 2237 to NGC 2244}

NGC 2237 was dubbed a twin cluster (or satellite cluster) to NGC 2244
by \citet{Li05}, as the stellar density distribution of 2MASS sources
shows a distinctive core to the west of the compact core of NGC 2244
(see Figure~1 in Li 2005). \citet{RomanZuniga08} also identified the
cluster, and pointed out that the region was historically recognized
as NGC 2237.  The region was not covered in the \citet{Poulton08} {\em
  Spitzer} survey.  Notably the NIR-excess disk fraction is
significantly lower than those in embedded clusters \citep[see][and
  Paper II]{RomanZuniga08}, but similar to that observed in the NGC
2244 cluster.

The NGC~2237 satellite cluster is very similar in various properties
-- obscuration, relative location and disk fraction to the NGC 2244
cluster -- to the X-ray-discovered RMC~XA cluster on the eastern side
of the Rosette Nebula (Paper II).  The separation between NGC 2237 and
the NGC 2244 cluster core is $\sim$14\arcmin\/ ($\sim$6 pc), almost
identical to the distance that separates RMC XA from NGC 2244. It is
plausible that pre-existing massive molecular clumps in the natal
cloud that formed NGC 2244 also collapsed and formed stars in RMC XA
and NGC 2237 simultaneously with the formation of NGC 2244. However,
they could also be triggered to form shortly following the onset of
the HII region around the NGC 2244 O stars.  Either way, these three
clusters represent the early star formation episode in the Rosette
complex.  The stars in the three clusters are probably
indistinguishable in age, and they share a similar low circumstellar
disk fraction (\S~\ref{nir_disk}).  NGC 2244 clearly has the dominant
OB population; NGC 2237 has many lower-mass stars but few OB stars,
and RMC XA is less rich. All three clusters show substructure: a
subcluster around the O5 star HD 46150 in NGC 2244 (Paper I); XA1, XA2, and
XA3 in RMC XA (Paper II); and elongated structures in NGC 2237
(\S~\ref{sec:concentration}).

\begin{figure}[H]
\centering
\plotone{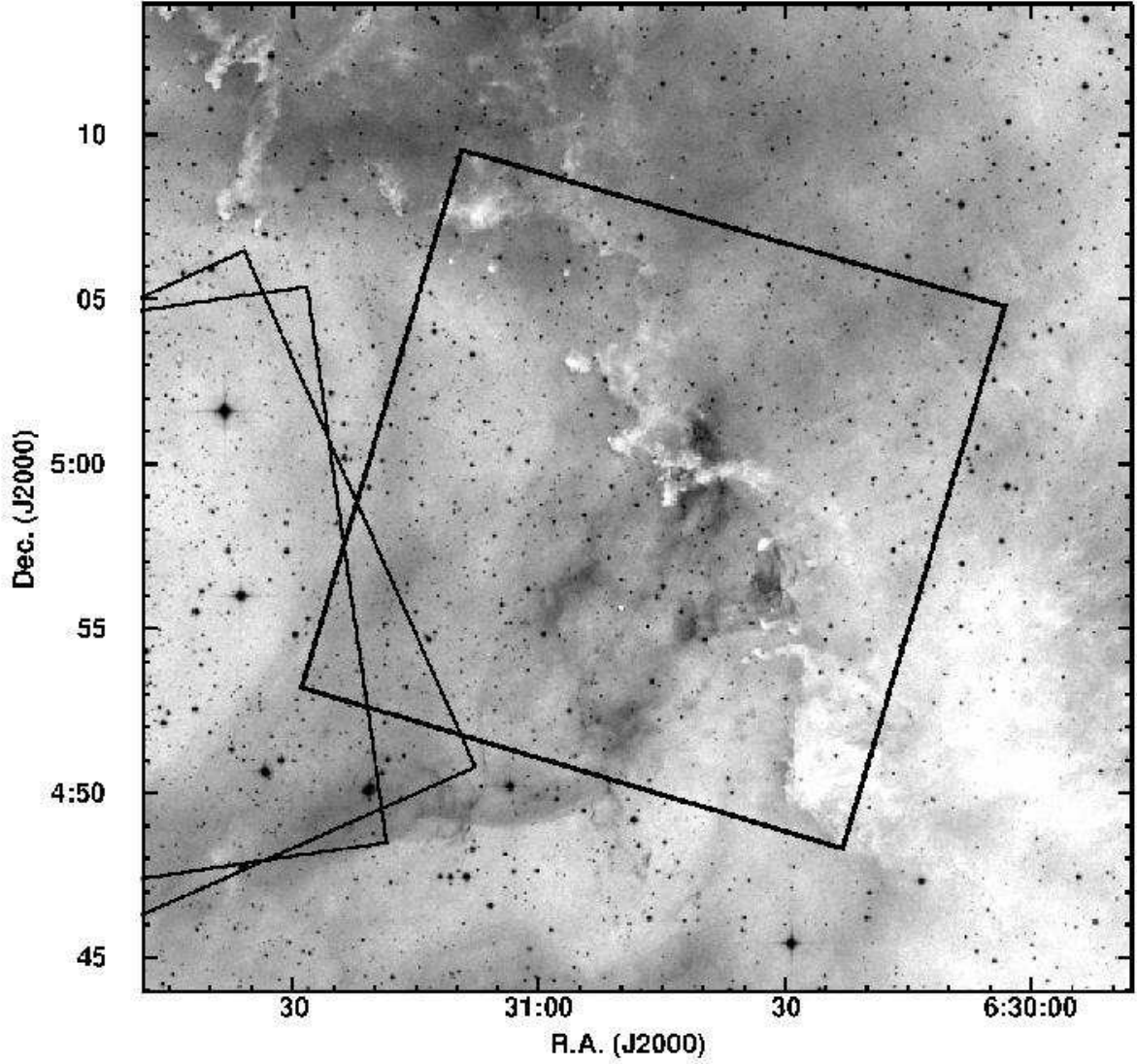}
\caption{A $30^{\prime} \times 30^{\prime}$ DSS2 $R$-band image of the
  NGC 2237 region. The box outlines the $17^{\prime} \times
  17^{\prime}$ ACIS-I field of view centered on NGC 2237. The fields
  to the east outline the deeper exposures of the rich NGC 2244
  cluster at the center of the Rosette Nebula presented in Paper I.\label{fig:dss}}
\end{figure}

\begin{figure}[H]
\plotone{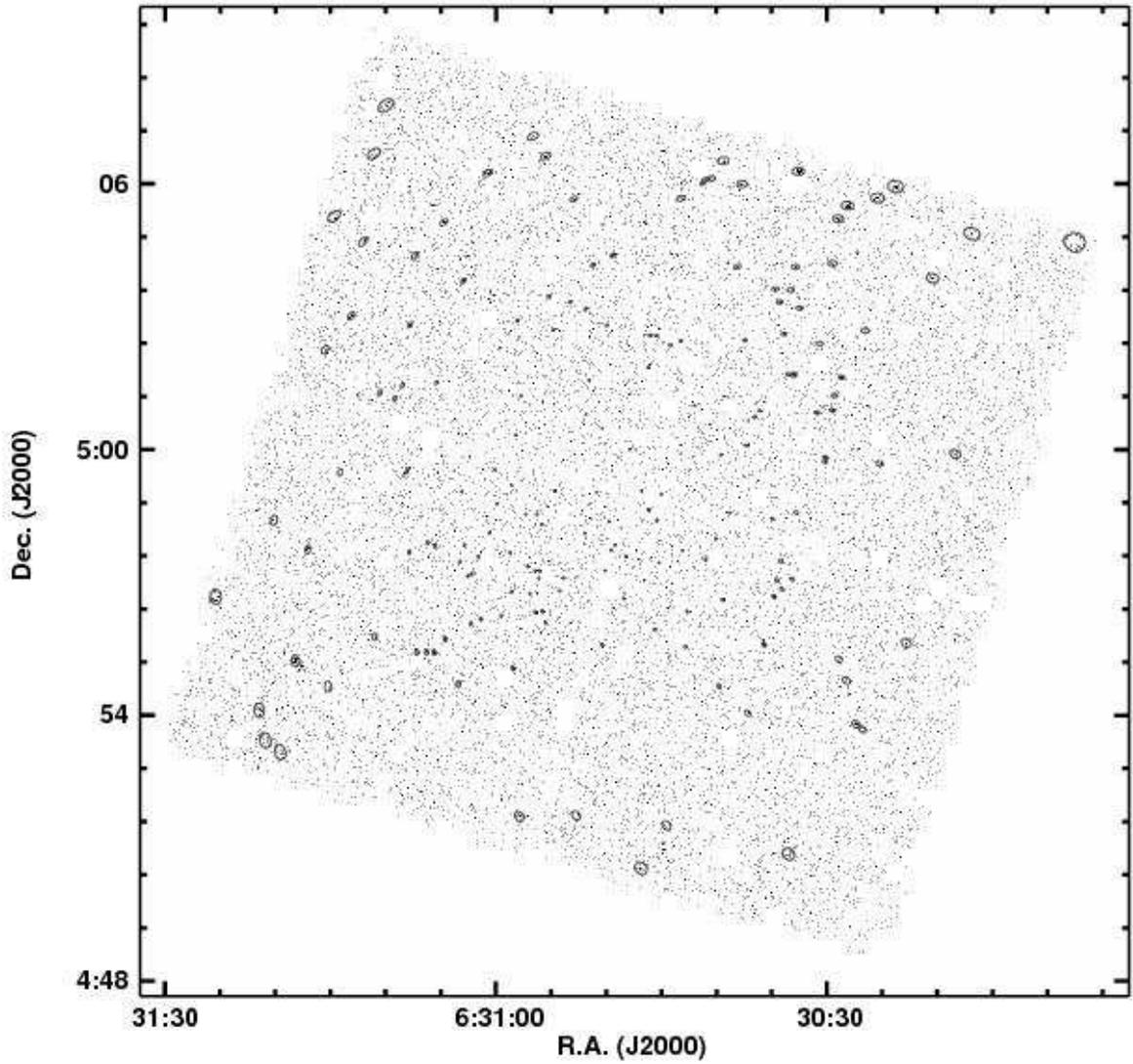}
\caption{$Chandra$ ACIS-I image of the NGC 2237 region (0.5-8 keV)
  overlaid with 168 source extraction regions.  The source extraction
  region is based on the local PSF, enclosing approximately 90\% of the PSF
  at 1.49 keV.\label{fig:acis}}
\end{figure}

\begin{figure}[H]
\plotone{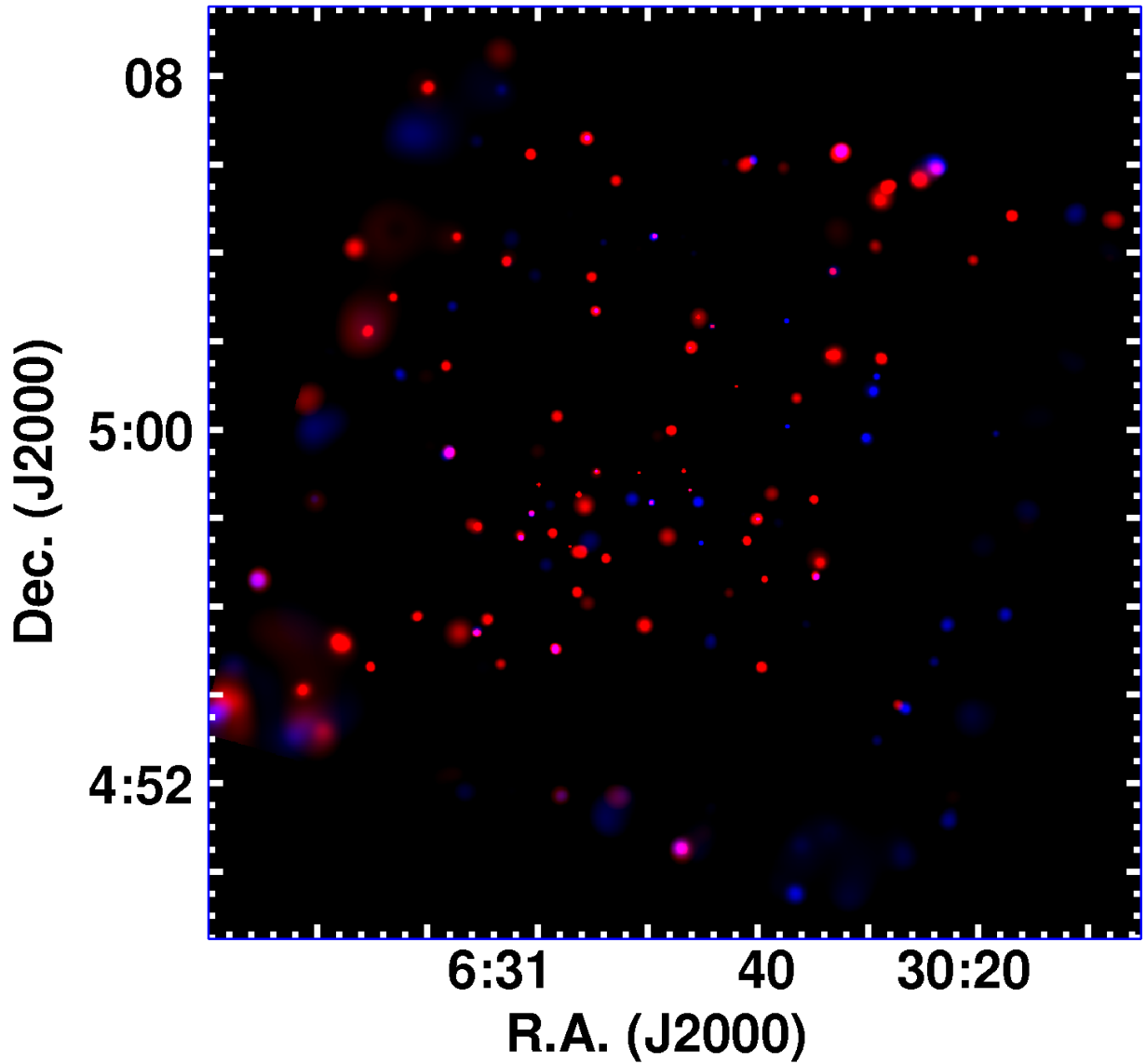}
\caption{Adaptively smoothed, exposure corrected ACIS-I image of the
  NGC 2237 region where red represents soft X-ray emission (0.5--2
  keV) and blue represents hard X-ray emission (2--7 keV).  Red
  sources are stars with emission dominated by soft counts.  Sources
  near the field edge appear larger due to optical distortion of the
  $Chandra$ mirrors.  \label{fig:csmooth}}
\end{figure}

\begin{figure}[H]
\epsscale{0.8}
\plotone{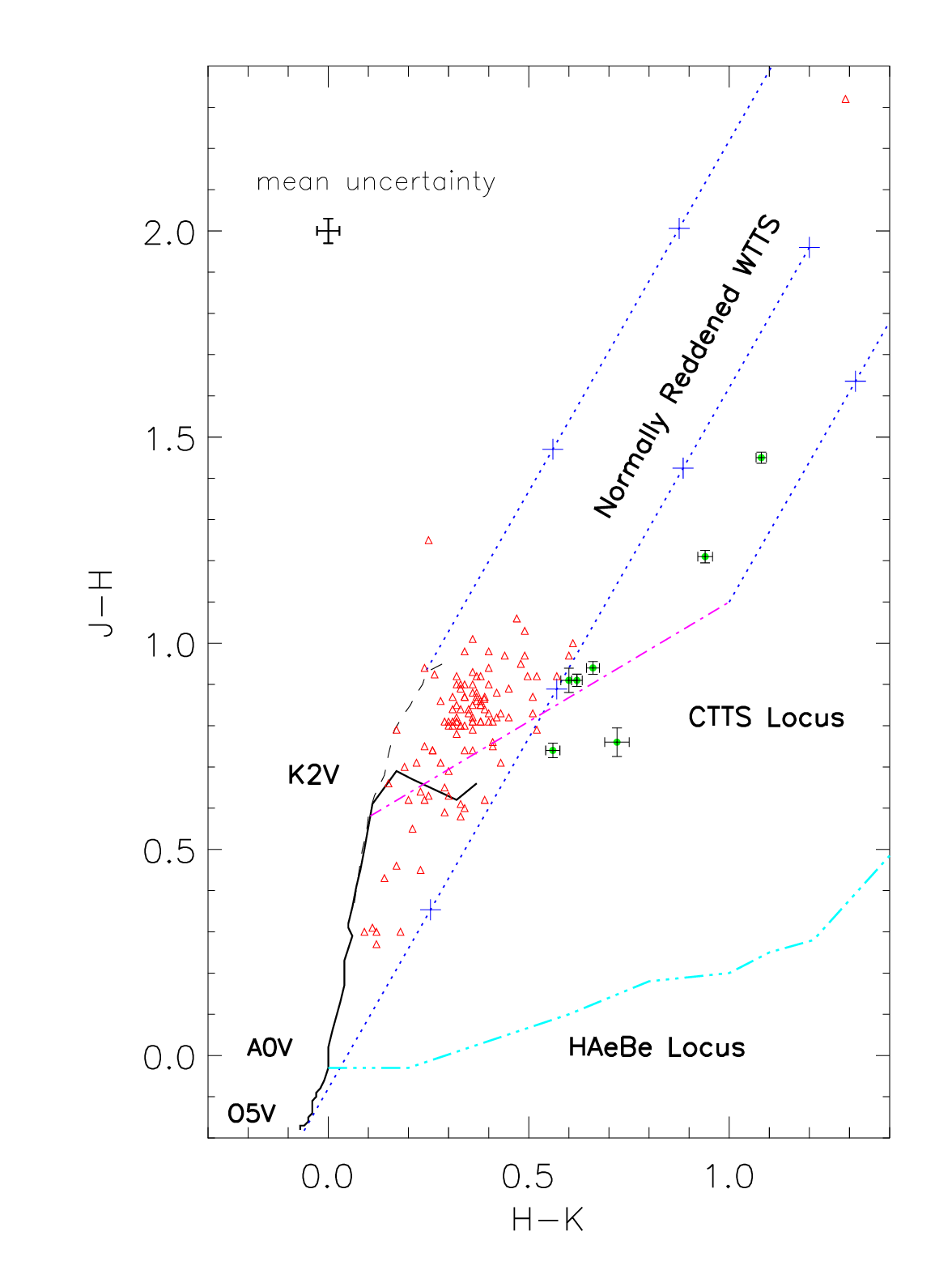}
\caption{NIR $J-H$ vs.\ $H-K$ color-color diagram for {\em Chandra}
  stars with high-quality $JHK$ photometry. The green circles and red
  triangles represent sources with and without significant $K$-band
  excess. The black solid and long-dash lines denote the loci of MS
  stars and giants, respectively, from Bessell \& Brett (1988). The
  purple dash dotted line is the locus for classical T Tauri stars
  from Meyer et al. (1997), and the cyan solid line is the locus for
  Herbig Ae/Be stars from Lada \& Adams (1992). The blue dashed lines
  represent the standard reddening vector with crosses marking every
  $A_V=5$ mag. The weak-line T Tauri stars are located in the domain
  between the blue dashed lines indicated as the WTTS
  locus.\label{fig:ccd}}
\end{figure}

\begin{figure}[H]
\plotone{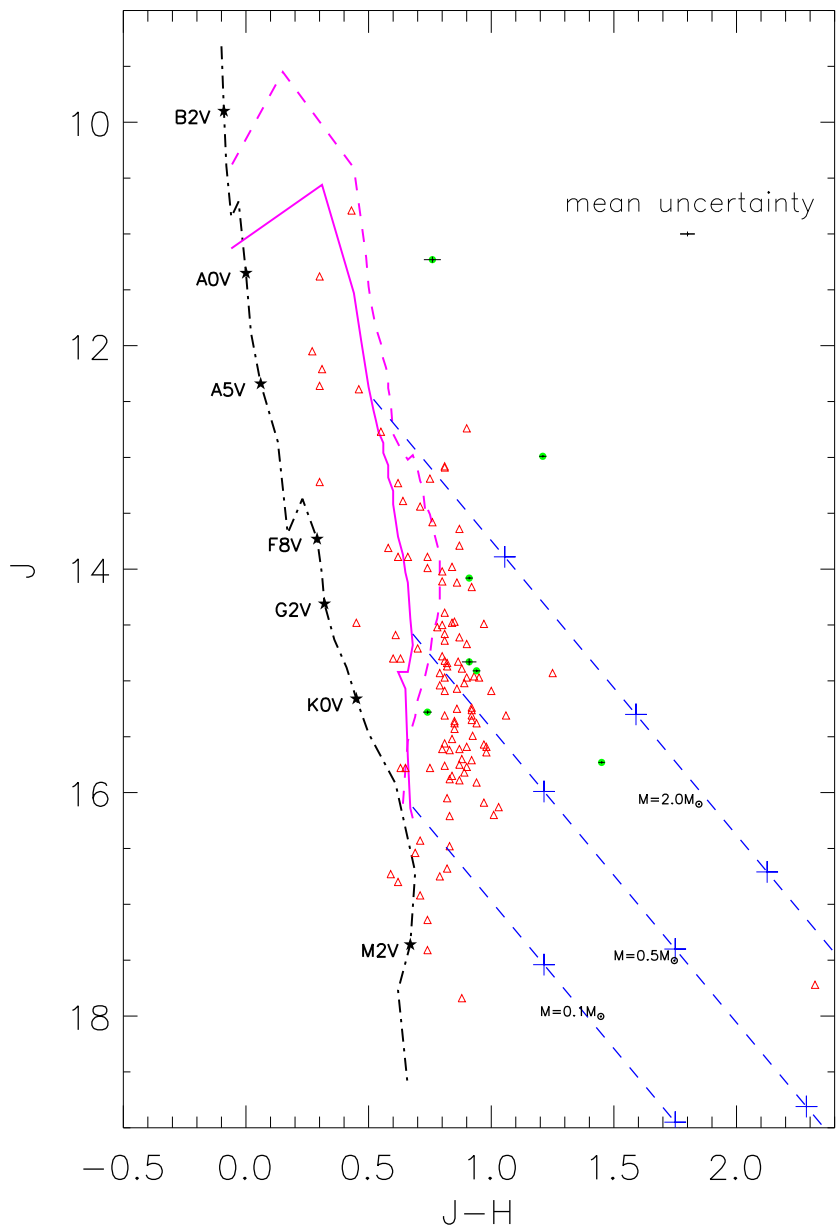}
\caption[NIR color-magnitude diagram for NGC 2237 {\em Chandra}
stars]{NIR $J$ vs. $J-H$ color-magnitude diagram using the same sample
and symbols as Figure~\ref{fig:ccd}. The purple solid line and dashed
line are the 2~Myr isochrone and the 1~Myr isochrone for PMS stars from
\citet{Siess00}, respectively. The dash dotted line marks the location
of Zero Age Main Sequence (ZAMS) stars. The blue dashed lines
represent the standard reddening vector with crosses marking every
$A_V=5$ mag and the corresponding stellar masses are
marked.\label{fig:cmd}}

\end{figure}

\begin{figure}[H]
\plotone{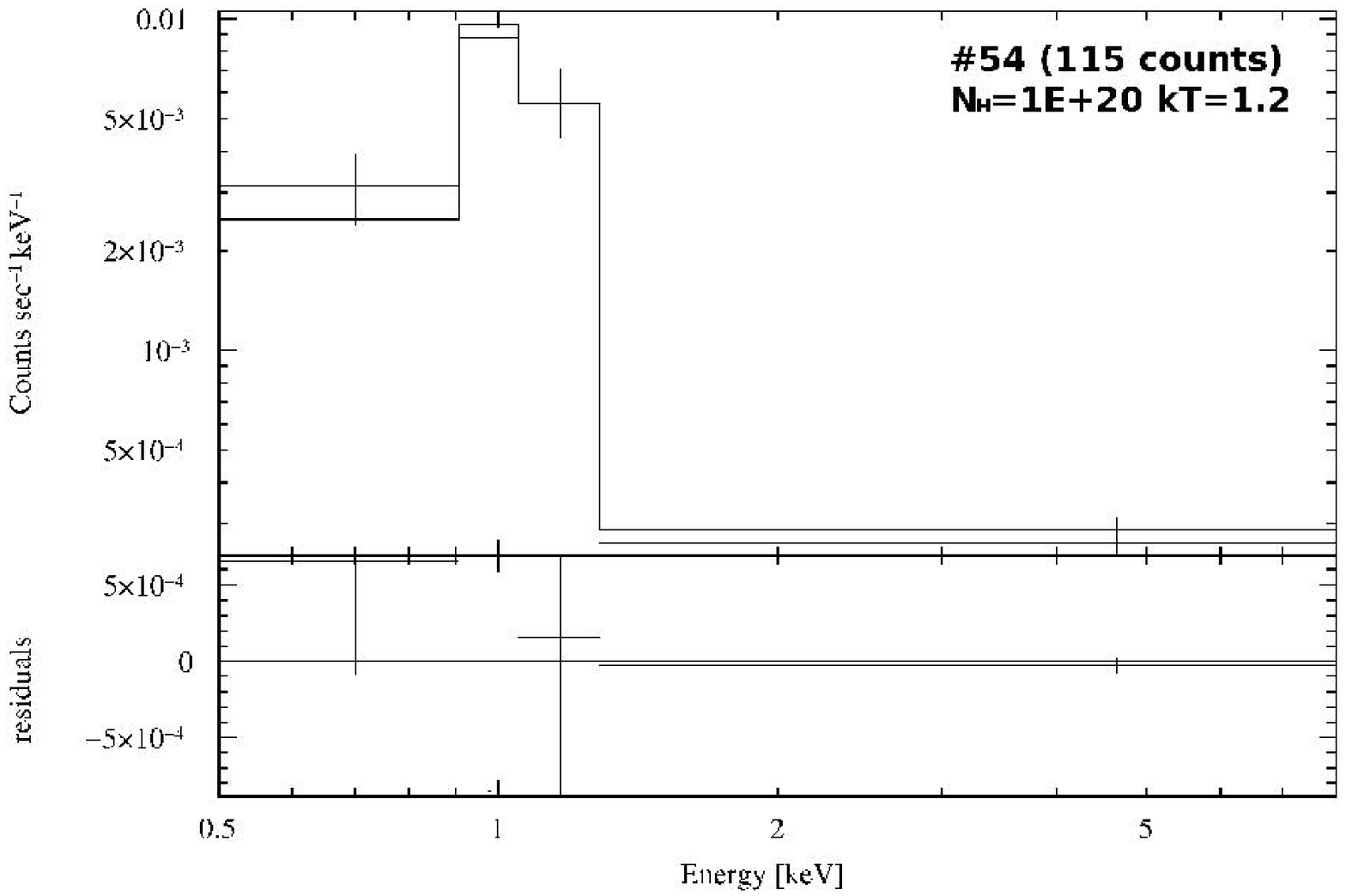}
\plotone{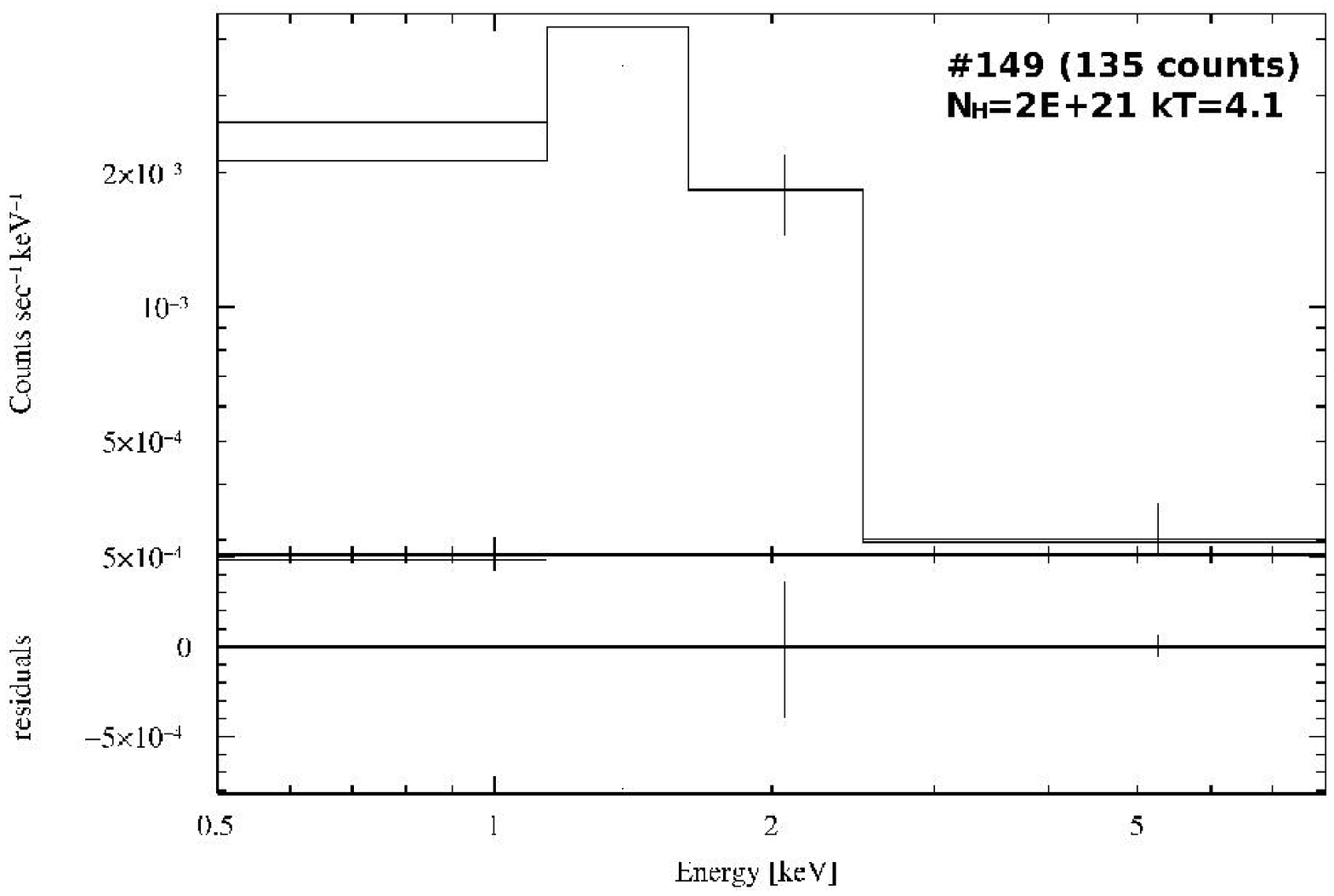}
\epsscale{1.0}
\caption{ (a) The X-ray spectrum and the spectral fit for the second
  brightest X-ray source in the field (\#54). (b) The X-ray spectrum
  and the spectral fit for the brightest X-ray star (\#149) in the
  field. See text for the best-fit model parameters.\label{fig:spec}}
\end{figure}

\begin{figure}[H]
\plotone{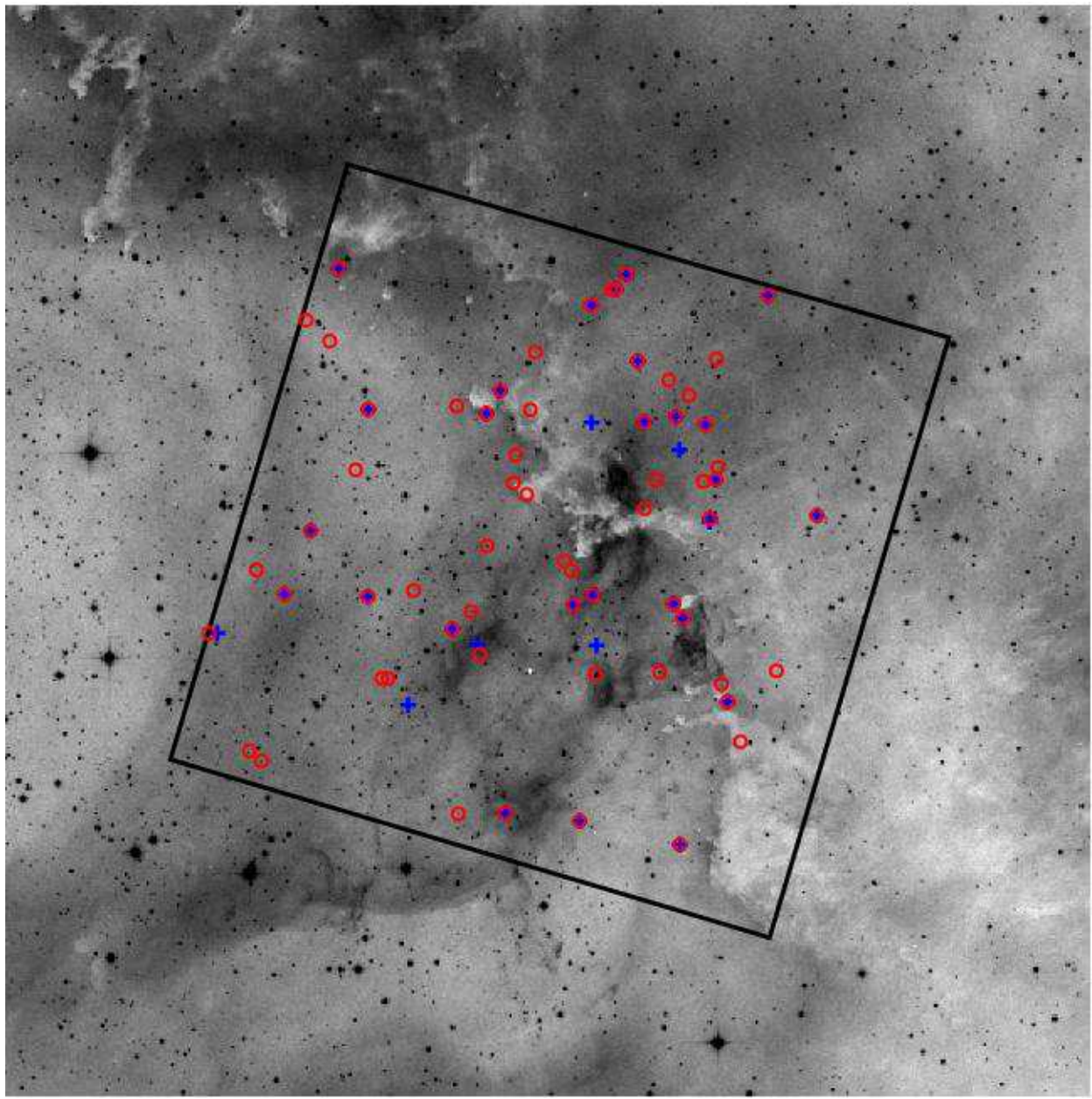}
\caption{The spatial distribution of X-ray sources that do not have
  matched IR counterparts (shown as blue crosses) and sources that
  have hard median photon energies ($MedE>2.0$ keV; shown as red
  circles). About ten hard X-ray sources without matched NIR
  counterparts are located inside the HII region cavity; these are
  likely AGNs. The rest, located inside the optical dark pillar region, are probably embedded young stars. \label{fig:hard}}
\end{figure}

\begin{figure}[H]
\plotone{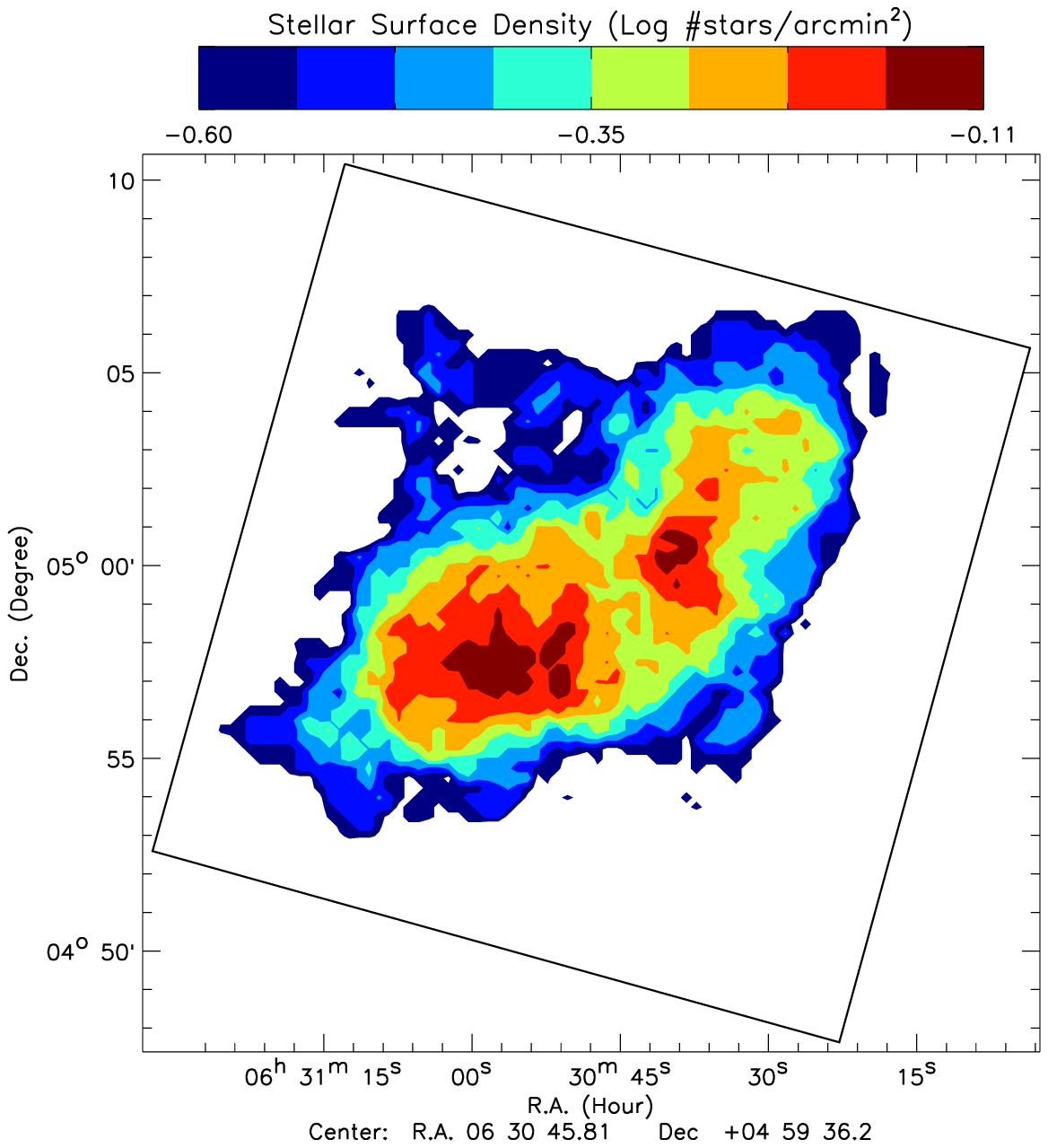}
\caption{Map of the stellar surface density, in units of $\log$\ stars
  per arcmin$^2$, for NGC 2237 sources with 6 or more net counts. It
  is smoothed with a 1.5\arcmin\/ radius sampling
  kernel.\label{fig:ssd}}
\end{figure}

\begin{figure}[H]
\plotone{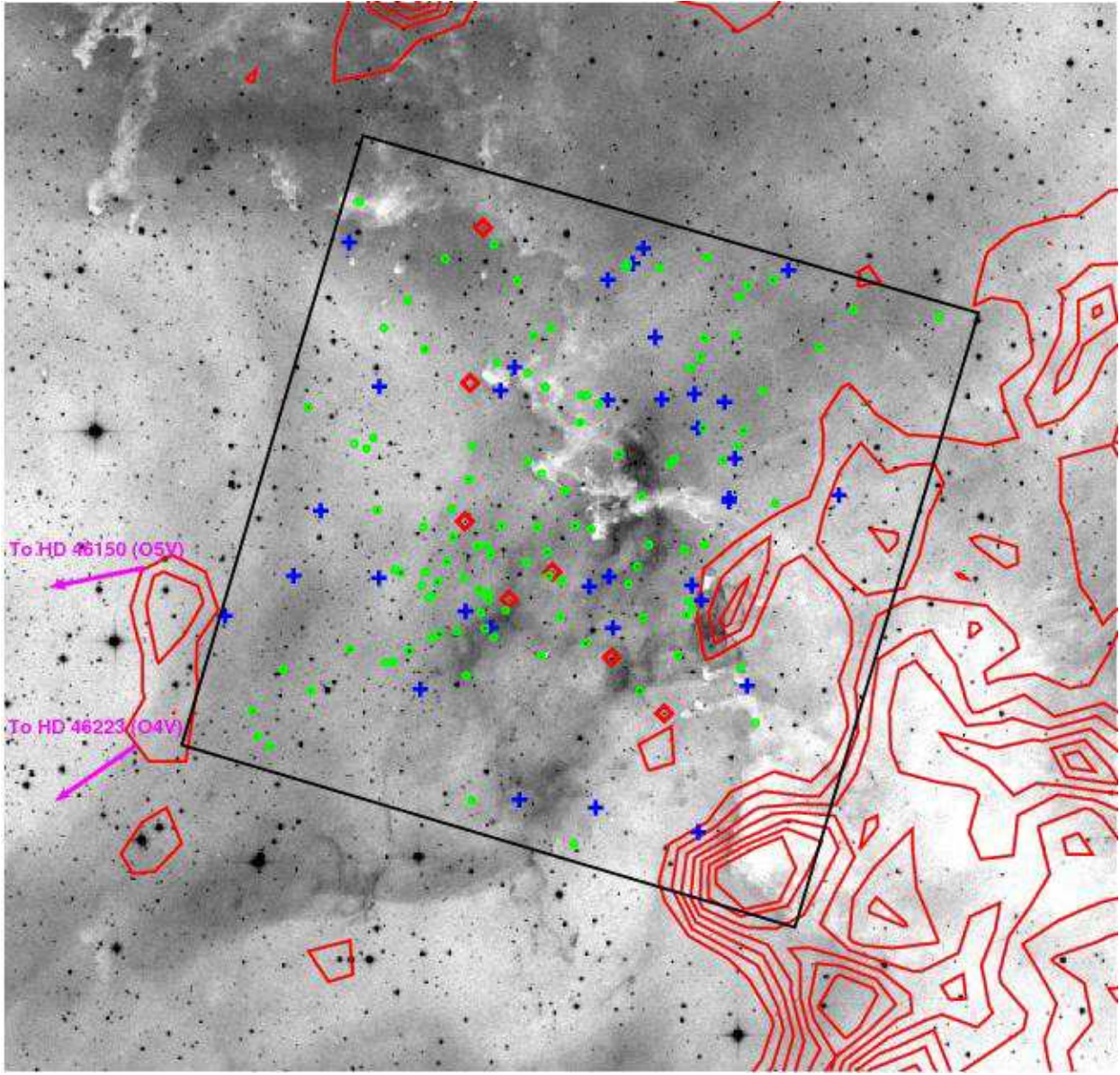}
\caption{The spatial distribution of the X-ray sources superposed on
  the optical DSS image. The X-ray-selected stars are Class II (red
  diamonds) and Class III (green circles) based on their NIR colors.
  Some X-ray sources without NIR counterparts (blue plusses) are
  likely embedded stars.  The molecular gas is outlined by the
  contours (red) from the $^{12}$CO emission survey of
  \citet{Heyer06}.  \label{fig:spatial}}
\end{figure}

\begin{figure}[H]
\plotone{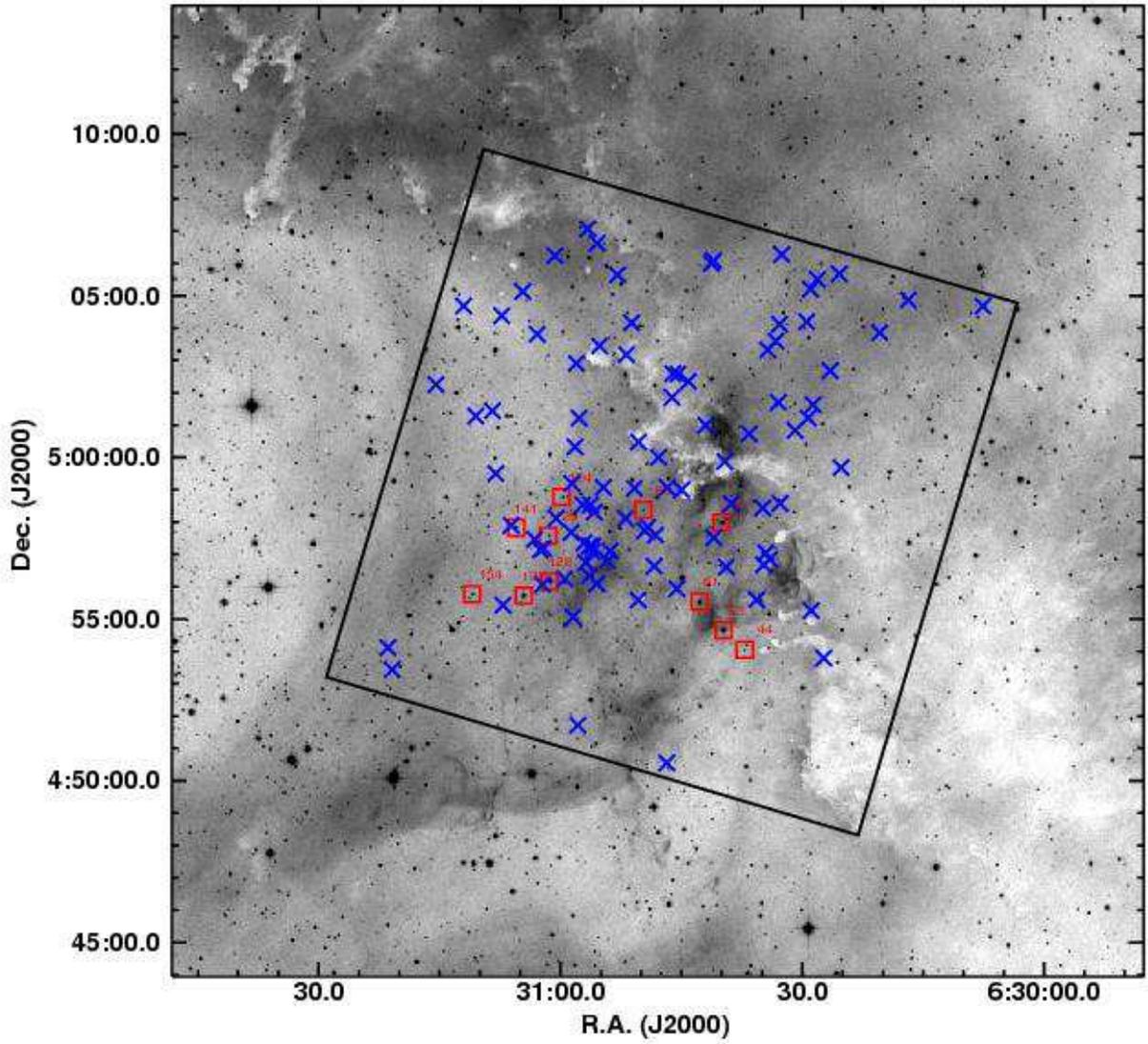}
\caption{Mass-stratified spatial distribution of NGC 2237 stars: $M\ga
  2\msun$ (red boxes), and $M< 2\msun$ (blue crosses). Masses are
  based on identified spectral types or estimated from NIR
  photometry.\label{fig:mass}}

\end{figure}

\begin{figure}
\plotone{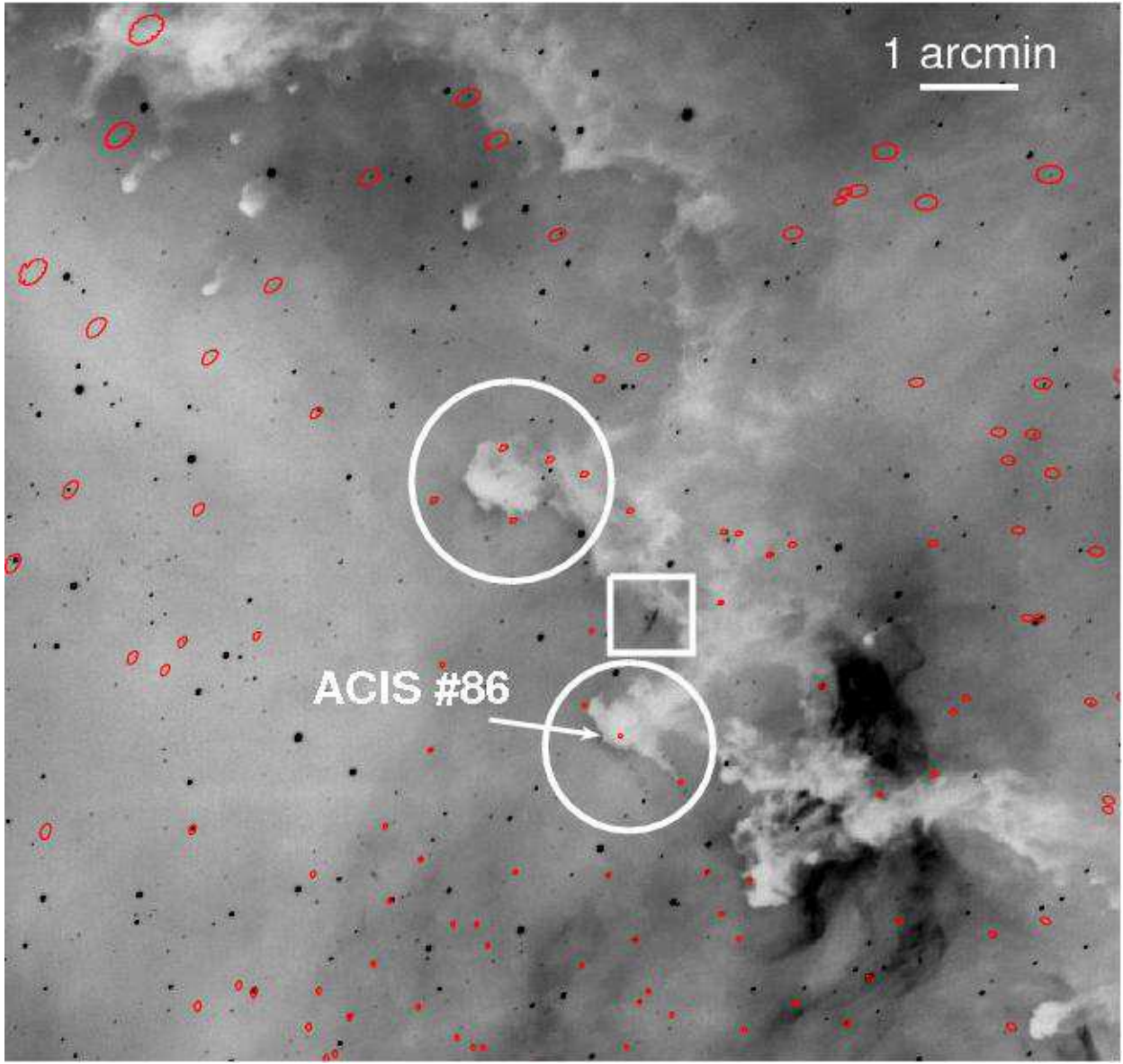}
\caption{The KPNO H$\alpha$ image of a region on the interface between
  the cloud and ionized nebula. Extraction regions of X-ray sources
  are shown in red. Two optically dark pillar objects are
  circled. Source \#86 (only 4 net counts) appears highly obscured
  with a visual extinction of $\sim 20$ mag in the NIR diagrams,
  possibly a star embedded in or behind the dense pillar. A likely
  Herbig-Haro object is marked with a box. \label{fig:halpha}}
\end{figure}

\begin{figure}[H]
\centering
\epsscale{0.7}
\plotone{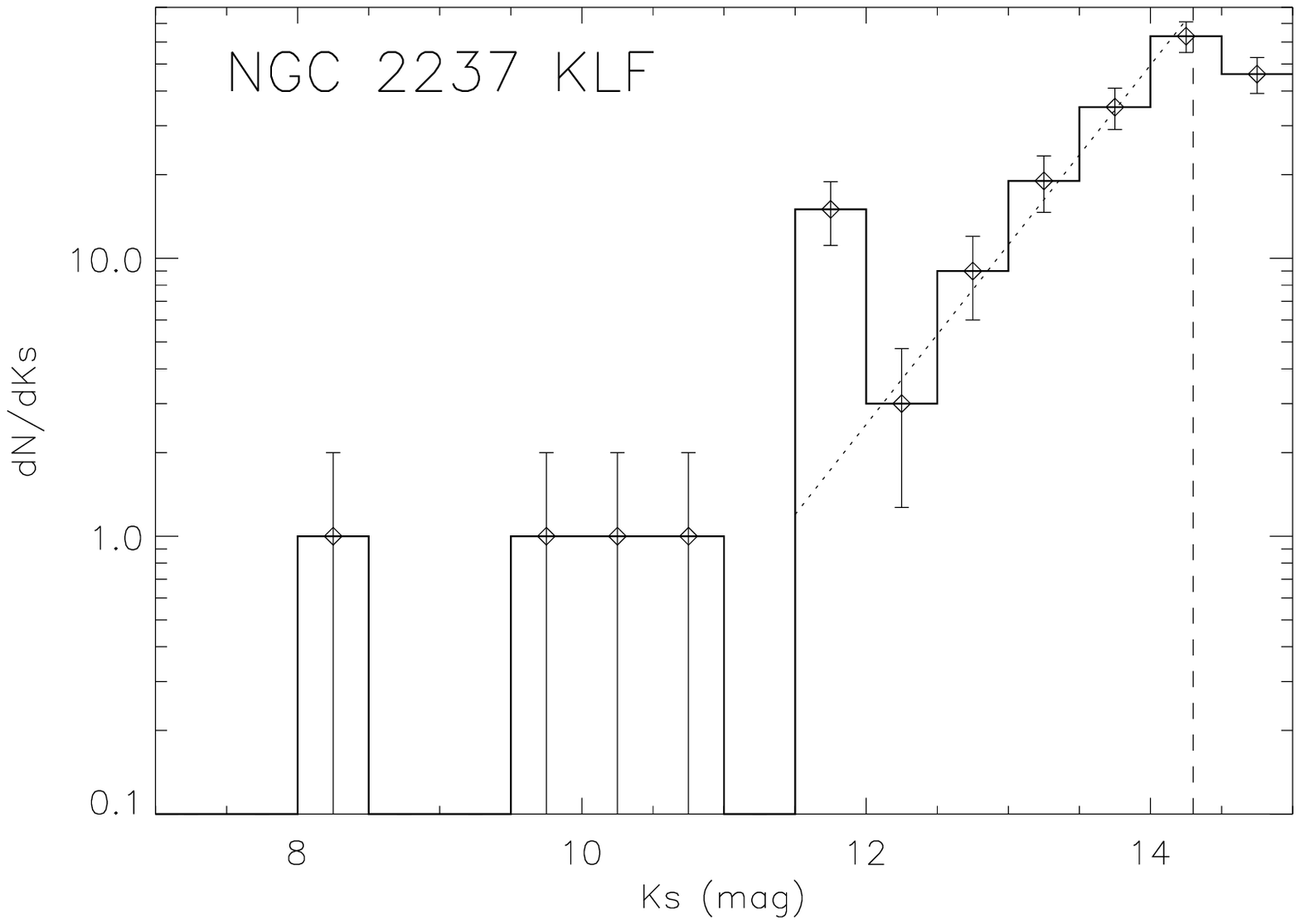}
\plotone{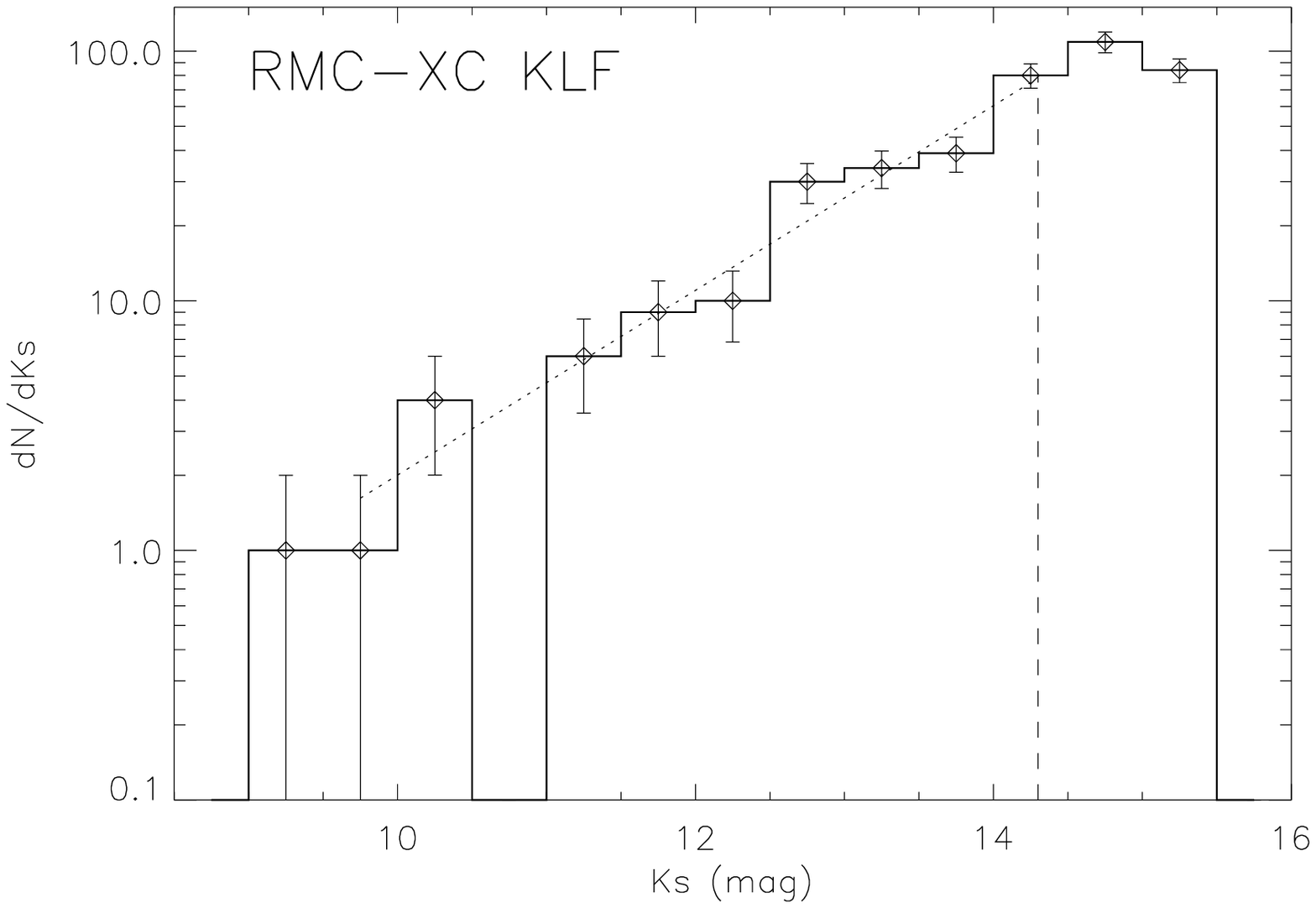}
\epsscale{1.0}
\caption{(a) The $K_s$-band luminosity function of NGC 2237 after
  correction for reddening (assuming $A_V=2.5$ mag) and background
  contamination. The KLF is found to match a power-law distribution
  with a slope of 0.6 (dotted line). The vertical dashed line
  indicates the 2MASS $K_s$ band completeness limit. (b) The KLF for
  cluster RMC XC (assuming $A_V=8$ mag; Rom{\'a}n-Z{\'u}{\~n}iga
  et~al.2008a). The power-law slope here is 0.39. \label{klf}}
\end{figure}

\begin{figure}[H]
\centering
\plotone{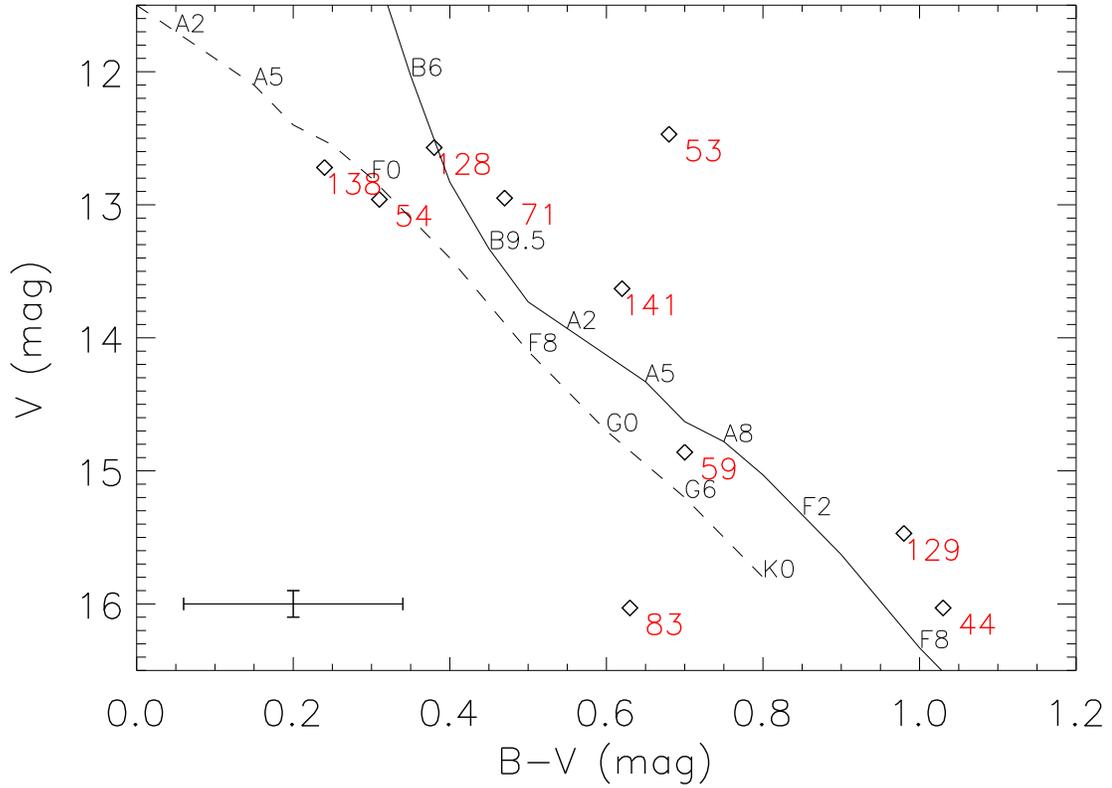}
\caption{The optical CMD for the 10 X-ray detected K-band bright
  sources, with ACIS source sequence numbers labeled. The solid line
  shows the zero-age main sequence (ZAMS) isochrone (Schmidt-Kaler
  1982) assuming a reddening of $E(B-V)$=0.5 and a distance of $d=1.4$
  kpc, whereas the dashed line represent the ZAMS location for
  foreground stars at 1 kpc with no reddening. The typical error bars
  for the photometry are shown.\label{10src}}
\end{figure}

\begin{figure}[H]
\plotone{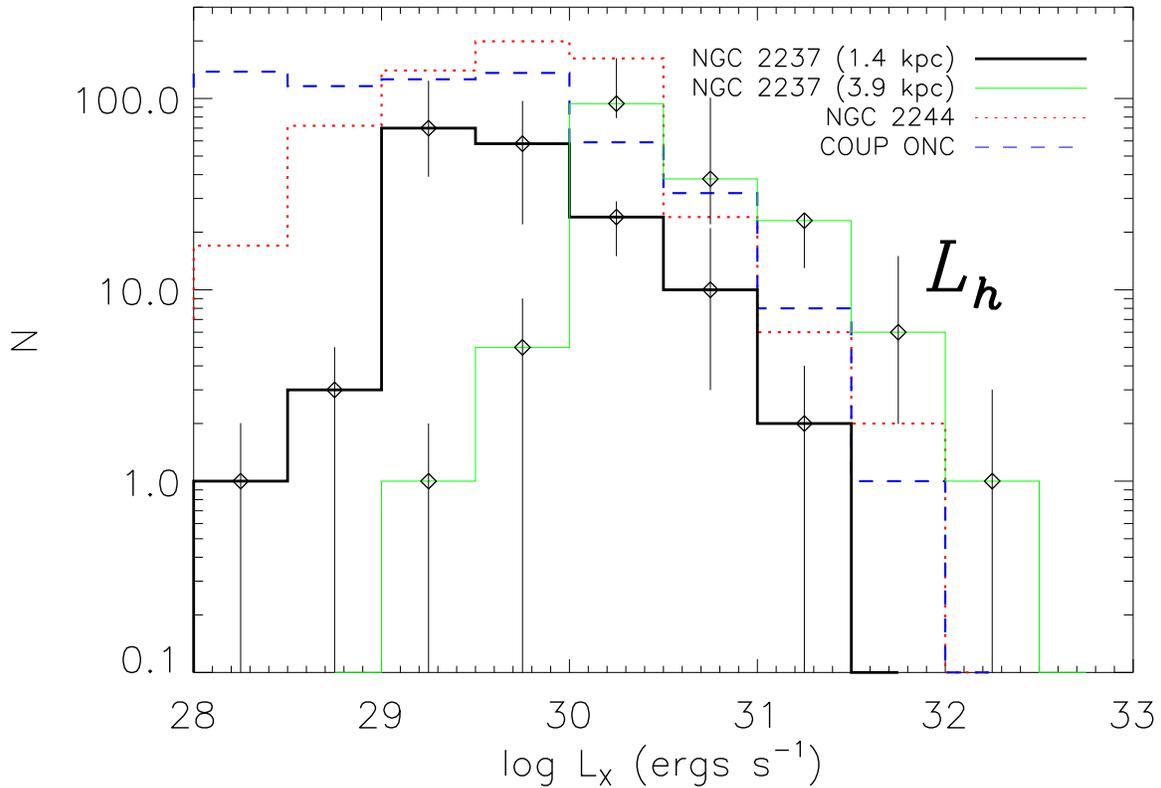}
\caption{The hard band (2-8 keV) X-ray luminosity function of
NGC 2237 (black solid line).  For comparison, the XLFs for NGC
2244 (red dotted line) and the COUP ONC (blue dashed line) are
plotted. The XLF derived using a farther distance $d=3.9$ kpc
(green solid line) is also shown. \label{xlf}}
\end{figure}

Further towards the west of the satellite cluster, where the molecular
pillars protrude into the HII region, the ionization front from the OB
stars in NGC 2244 may have triggered some star formation. These
``elephant trunk'' globules \citep{Schneps80} and bright rims (see
Figure~\ref{fig:dss} and optical images of the Eagle Nebula) are
commonly seen at the peripheries of many nebulae and are interpreted
as illustrations of the massive star feedback into the parent
molecular cloud.  The evidence for triggered star formation in
photoevaporating pillars is tantalizing. The alignment of the Class II
sources in NGC 2237 may be a hint of an ionization front that
triggered collapse along the boundary of the HII region
(\S~\ref{sec:concentration}), but the number of stars involved is
small. No clear evidence of an age gradient (or evolutionary sequence)
is found.  \citet{Dent09} have studied the nature and mass of the
small CO clumps in the Rosette including the elephant
trunks/globulettes in the NGC 2237 region.  They show that the
globulettes are mostly low mass, which are being probably accelerated
by the OB winds, and their possible age is shorter than 1 Myr.  The
removal of the rim gas proceeds quite rapidly, thus NGC 2237 must have
emerged relatively fast from its original clump.

Only a handful of new stars are forming at the vertices of the
elephant trunks in NGC 2237, perhaps fewer than the number of
protostars at the tip of molecular pillars in M16 \citep{Hester05}.
The masses of the clumps in the expanding ring range between
$\sim$3œôòó-100 $M_{\odot}$ \citep{Dent09}.  Star formation efficiency of
a few percent will suffice to form the stars observed in the mass
range of NGC 2237.  Dynamical instability \citep{Garcia-Segura96}
appears to be not effective in forming stars here. As shown in
\citet{Whalen08}, expanding HII regions can either stimulate or lead
to turbulent flows against the formation of later generations of
stars, depending on the efficiency of cooling within the shocked
material. The observed deficiency of stars forming at the ionization
front here may imply that the cooling is weak and dynamical
instability is not promoting formation of second-generation stars
efficiently.

Note however that the paucity of X-ray stars in the pillars could be
due to low sensitivity of our observation in this region.  The X-ray
luminosity of protostars may also be intrinsically low in the
evaporating gaseous globules (EGGs): \citet{Linsky07} report X-ray
non-detection of EGGs in M16 and conclude that either the EGGs do not
contain protostars or the protostars have not yet become X-ray active.

More X-ray sources are seen distributed ahead of the ionization front;
these account for the density enhancement northwest of NGC 2237 seen
in Figure~\ref{fig:ssd}. Most of these X-ray sources show hard X-ray
emission, suggesting that they are embedded young stars. The presence
of a likely HH object (\S~\ref{sec:concentration}) supports the
ongoing formation of stars in this region. These stars may have formed
as a result of interaction between the molecular materials here and
the Rosette Nebula. Spectroscopy and {\em Spitzer} mid-IR data can
determine their ages and evolutionary stages.

\subsection{Large Scale Star Formation in the Rosette Complex}

\begin{figure}[H]
\epsscale{0.8}
\plotone{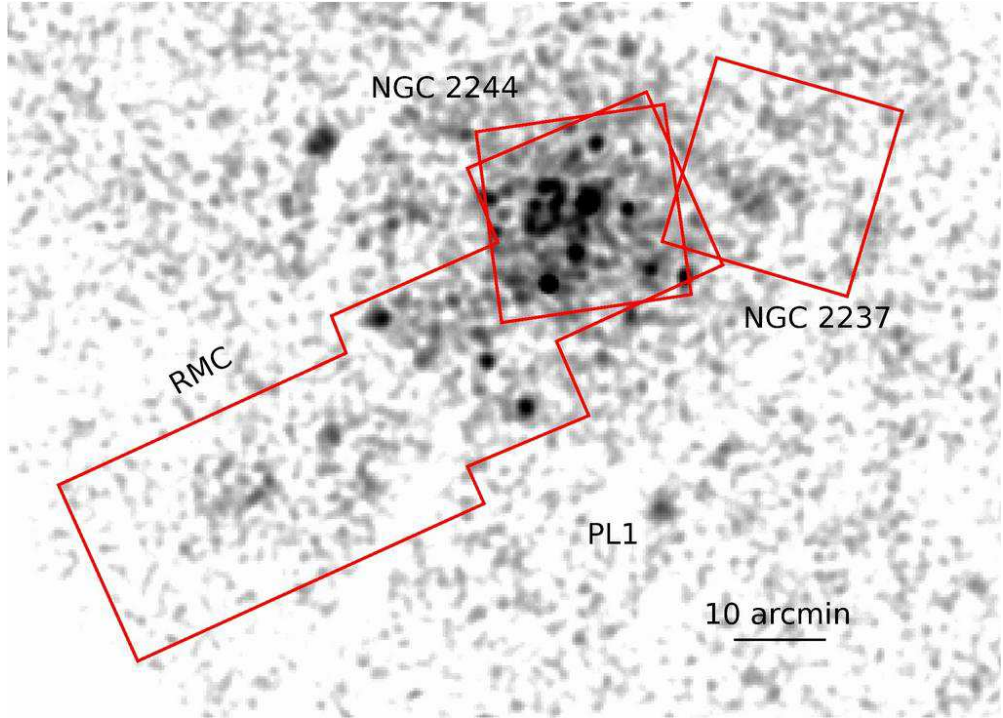}
\epsscale{1.0}
\caption{(a) $ROSAT$ PSPC image of the Rosette Complex with existing
  Chandra pointings outlined by the red boxes.  (b) {\em Chandra}
  ACIS-I mosaic of six observations in the Rosette Complex. Red
  represents soft band (0.5--2 keV) X-ray emission, and green represents
  hard band (2--7 keV) X-ray emission. The Chandra image is rotated
  for better layout in the page; the actual orientation is the same as
  (a), where north is up and east is to the left. A similar image
  without the NGC 2237 exposure is shown by
  \citet{TFM03}.\label{fig:mosaic1}}
\end{figure}

Figure~\ref{fig:mosaic1}a shows the $ROSAT$ PSPC image with complete
spatial coverage of the region revealing the brightest X-ray-emitting
stars.  For a high resolution comparison, we assemble our {\em
  Chandra} images (exposure corrected) in a $\sim 1.25^{\circ} \times
0.25^{\circ}$ mosaic, as shown in Figure~\ref{fig:mosaic1}b; a similar
mosaic was shown by \citet{TFM03}.  The clumps of stars are apparent,
which gives an overall view of their relative distributions, although
the detection sensitivity is not uniform.  To better understand the
star formation history of this complex, we need to examine the large
scale cluster formation events as an integrated process and look
beyond our X-ray coverage.

To outline the relative locations of the clusters with respect to the
molecular material and the morphology of the ISM,
Figure~\ref{fig:mosaic2}a shows the clusters superposed on the
$^{13}$CO emission map from \citet{Heyer06}.  The NGC 2244 cluster has
apparently expelled most of the molecular gas while creating the
central cavity. The RMC XA cluster and NGC 2237 are mostly unobscured
without surrounding molecular material, and the stars have low
NIR-excess disk fraction. They may have formed nearly simultaneously
or immediately after the formation of NGC 2244. The estimated total
population is not as rich as NGC 2244, on the order of 200-600
($\sim$10-30\% of the total stellar population in NGC 2244). However,
their stellar masses were probably determined by the available
molecular material in their parental clumps; no strong feedback that
may disrupt the cloud and suppress the star formation is expected from
the NGC 2244 OB stars yet as they were forming.

Moving further radially outwards, 25\arcmin--30\arcmin\/ ($10-12$ pc)
from the central OB cluster, a number of other clusters are present:
RMC XB from Paper II (PL2, cluster C in Poulton et al.\ 2008, or
structure B in Li \& Smith 2005) to the southeast, PL 1 (cluster A in
Poulton et al. 2008, also structure B in Li \& Smith 2005) to the
south, cluster REFL10 \citep{RomanZuniga08} to the north.  We also
noticed a significant overdensity of stars associated with the
molecular clump to the west of NGC 2237 in both the 2MASS study (see
Figure 5 in Li 2005) and the FLAMINGOS NIR survey \citep[Figure 27
  in][]{RomanZuniga08}.  The core is clearly seen in the CO map in
\citet{Heyer06} that shows prominent, dense molecular clumps further
westwards from NGC 2237 (approximate center
R.A.=$06^{\rm{h}}29^{\rm{m}}$, Dec.=$+05^{\circ}00^{\prime}$). This
cloud core was previously identified as one of the principal regions
in the RMC, designated as Core E
\citep{Blitz80}. \citet{RomanZuniga08} find a collection of highly
reddened NIR stars without $J$-band detections here. Neither {\em
  Chandra} nor {\em Spitzer} satellites have covered this molecular
core. Nevertheless, we speculate that it could well be another region that
harbors an undiscovered small embedded cluster.

As seen in Figure~\ref{fig:mosaic2}a, there is a prominent large
shell-like structure in the CO map, with the above clusters located
along the shell coinciding with the densest spots.  Presence of such
dense molecular material has been reported in other recent studies of
star formation triggered by the expansion of an HII region
\citep[e.g., ][]{Deharveng09,Kang09}.  The distribution of the
clusters is also simplified in the cartoon presentation in
Figure~\ref{fig:mosaic2}b.  These clusters could be formed through the
so-called ``collect-and-collapse'' process \citep[for review,
  see][]{Elmegreen98,Deharveng05} involving fragmentation of swept-up
shells of interstellar gas by the expanding HII region of NGC 2244. When
critical condensation is reached, the layers of neutral gas become
gravitationally unstable and eventually fragment to form stars
\citep{Whitworth94}.

To elucidate this collect-and-collapse model, we adopt typical values
in models for cloud fragmentation induced by expanding HII regions and
stellar wind bubbles \citep{Whitworth94}.  Assuming an isothermal
sound speed $a_s=0.2$~km s$^{-1}$ in the shocked gas, an emitting rate
of hydrogen ionizing photons $\dot{N}_{LyC}\sim 10^{49}$ s$^{-1}$ for
the O4 and O5 stars, a wind luminosity $L_w\sim 10^{37}$ erg s$^{-1}$,
and a hydrogen nuclei number density $n_0\sim 1000$ cm$^{-3}$ in the
interstellar medium, the timescale at which fragmentation occurs is
$\sim 1-2$ Myr and fragmentation starts in the shell at radii of
$6-10$ pc with fragment masses around $10-23$~M$_{\odot}$. The ring of
clusters around $10-12$ pc (RMC XB = PL 2, PL 1, and REFL 10)
correspond to this predicted ring of collapsing cloud fragments. Note
the presence of a massive Class I protostar ($Chandra$ source \#89 =
IRAS 06306+0437 in Paper II) in a sparse cluster within RMC XB.

Several observations might confirm this collect-and-collapse
scenario. The model predicts that active star formation, small
clusters with high disk fractions, would be present in PL 1 and REFL
10 on the CO rim. The model also suggests that PL 2 formed more
recently than more distant clusters in the RMC. Consistent with this
prediction, \citet{RomanZuniga08} found that the IR excess fraction in PL 2
($33\% \pm 6\%$) is higher than PL 4 ($24\% \pm 3\%$). The
collect-and-collapse process that formed PL2 may be enhanced from the
expansion of cluster PL4 and PL5.

We note that the four cloud fragments around the central NGC 2244
cluster form a bowl that is open towards the northeast, without
significant molecular material or clusters. In this direction lies the
large Monoceros Loop supernova remnant \citep[SNR;][]{Fountain79,
  Vidal82, Odegard86}, and one may speculate that it may have some
influence to the RMC.  However, \citet{TFM03} reviewed the possibility
of an encounter between the SNR and the molecular cloud, and there is
no sign of interaction.

Farther away from NGC 2244, the very young clusters PL4, PL5, REFL08
(together they are recognized as cluster RMC XC in Paper II, as
structure C in Li \& Smith 2005, and as cluster E in Poulton et
al. 2008), PL6, PL7 (cluster G in Poulton et al. 2008), and REFL09
(cluster F in Poulton et al. 2008) are located along the densest
mid-plane of the cloud.  This was previously taken as evidence of
sequentially triggered star formation \citep{Elmegreen77}, where one
cluster provokes the formation of the next one.  Based on the NIR
studies of the embedded clusters, \citet{RomanZuniga08} propose a
different idea of sequential formation. The IR-excess disk fraction
among the clusters increases away from the nebula suggesting a rough
sequence of ages, but 2 Myrs may not be enough time to produce a
triggered sequence. The clusters PL7 and REFL09 are too close in age
and distance to be triggered by clusters PL4 and PL5. Their formation
is likely an outcome of the evolution of the primordial molecular
cloud, and not from the interaction with the HII region
\citep{RomanZuniga08}.  In fact, recently \citet{Williams09} show that
star formation in PL06 must have been active for about 2 Myrs and
still continues.  This reinforces the view that cluster formation in
the RMC may have proceeded without significant time difference, except
that NGC 2244 has more OB stars with powerful winds and clears its
surroundings more rapidly than the other low mass star clusters (e.g.,
NGC 2237, REFL 10) and deeply embedded small clusters (e.g., PL 06).

\section{Conclusions \label{sec:summary.sec}}

We present the first high spatial resolution X-ray image of the NGC
2237 cluster in the Rosette Nebula obtained via a single {\it Chandra}
observation, which is part of our survey of the star formation
activities in the Rosette complex.  Prior to this study, only 36
$K$-band excess cluster members were identified \citep{RomanZuniga08}.
The effectiveness of the X-ray survey technique is clearly
demonstrated.  Our main findings are:

1.  In this 20 ks observation, 168 X-ray point sources are detected
with a limiting X-ray sensitivity of $L_t\sim 10^{30}$ ergs s$^{-1}$,
nearly complete to the solar mass range.  We associate 134 of the 168
X-ray sources with cluster stars and infer about 24 other embedded
sources, giving a total of $\sim 160$ stellar members identified
here in NGC~2237.

2. The locations of most ACIS sources in the NIR color-magnitude diagram
are consistent with a small population of PMS low mass stars ($M \la 2
M_{\odot}$) in the NGC 2237 region with a visual extinction of $A_V
\sim 2$ at 1.4 kpc, assuming a similar age to NGC 2244 (2 Myr old).
Our X-ray sample provides the first probe of the low mass population
in this satellite cluster to NGC 2244.  We derive an overall
$K$-excess disk frequency of 13\% for stars with mass $M \ga
1M_{\odot}$ using the X-ray-selected sample, consistent with the
reported disk fraction from NIR study.

3.  We examine the X-ray properties of a number of B stars in the
field. Several X-ray emitting stars are located at the peripheries of
pillar objects. A previously unknown optical outflow feature is seen
originating inside the optically dark region, which may be a newly
identified HH object; this supports the ongoing formation of stars in
this region.

4. The X-ray selected population provides a reliable probe of the
cluster center and structure, which agree well with previous NIR
studies. Similar to RMC XA, collapse of pre-existing massive molecular
clumps accompanying the formation of the NGC 2244 cluster may have
formed NGC 2237 as a satellite cluster.  There is tantalizing evidence
suggestive of a triggered formation process in the optical dark
pillars by the NGC 2244 O stars, including discovery of young Class II
sources aligned in an arc and embedded hard X-ray sources across the
optical Photon Dominated Region.

5.  We derive the KLF and the XLF for the NGC 2237 cluster.  A
significant excess of stars with apparent $K_s\sim 11-12$ is seen in
the KLF, likely associated with the presence of a few foreground stars.
The slope of the NGC 2237 XLF matches the ONC XLF well, and the
relative scaling implies a smaller cluster population $\sim$400--600
for NGC 2237, which is consistent with the estimates from NIR studies.
We discuss the uncertainty in the distance to NGC 2237 and prefer the
near distance.

6.  The large-scale star formation in the Rosette Complex is reviewed
in the context of the ``collect-and-collapse'' scenario, given the
locations of clusters and the morphology of the ISM.  The formation of
NGC 2244, RMC XA, and NGC 2237 represents the earliest star formation
episode in the complex.  The IRAS 100$\mu$m image and CO emission map
show sites of currently active star formation--four potential clusters
associated with the fragments of the large shell of swept-up neutral
material from the expanding HII region.  In the RMC, a temporal
sequence of star formation across the complex is present \citep[see
][]{Li05c,Wang08b,RomanZuniga08}.  The cause of the sequence appears
to be related to the molecular cloud, perhaps primordial, not limited
to the originally proposed triggering scenario in \citet{Elmegreen77}
where one cluster provokes the formation of the next one.

We thank the anonymous referee for his/her careful reading and
comments that significantly improved the clarity of this paper.  We
thank Travis Rector and Mark Heyer for kindly providing the KPNO
MOSAIC images of the Rosette Nebula and the CO emission maps of the
Rosette Complex, respectively.  This work was supported by the Chandra
ACIS Team (G. Garmire, PI) through NASA contract NAS8-38252.  E.D.F.,
P.S.B., and L.K.T. also received support from NASA grant
NNX09AC74G. FLAMINGOS was designed and constructed by the IR
instrumentation group (PI: R. Elston) at the University of Florida,
Department of Astronomy with support from NSF grant AST97-31180 and
Kitt Peak National Observatory.  The data were collected under the
NOAO Survey Program, ``Towards a Complete Near-Infrared Spectroscopic
Survey of Giant Molecular Clouds'' (PI: E. Lada) and supported by NSF
grants AST97-3367 and AST02-02976 to the University of
Florida. E.A.L. also acknowledges support from NASA LTSA NNG05D66G.
This publication makes use of data products from the Two Micron All
Sky Survey (a joint project of the University of Massachusetts and the
Infrared Processing and Analysis Center/California Institute of
Technology, funded by NASA and NSF), and the SIMBAD database and the
VizieR catalogue access tool (operated by the CDS).


\clearpage


\begin{deluxetable}{rcrrrrrrrrrrrccrr}
\centering \rotate \tabletypesize{\scriptsize} \tablewidth{0pt}
\tablecolumns{17}

\tablecaption{Primary {\em Chandra} Catalog:  X-ray Source Properties \label{tbl:primary}}

\tablehead{
\multicolumn{2}{c}{Source} &
\multicolumn{4}{c}{Position} &
\multicolumn{5}{c}{Extraction} &
\multicolumn{6}{c}{Characteristics} \\

\multicolumn{2}{c}{\hrulefill} &
\multicolumn{4}{c}{\hrulefill} &
\multicolumn{5}{c}{\hrulefill} &
\multicolumn{6}{c}{\hrulefill} \\

\colhead{Seq} & \colhead{CXOU J} &
\colhead{$\alpha$} & \colhead{$\delta$} & \colhead{$\sigma$} & \colhead{$\theta$} &
\colhead{$C_{t}$} & \colhead{$\sigma_{t}$} & \colhead{$B_{t}$} & \colhead{$C_{h}$} & \colhead{PSF} &
\colhead{Signif} & \colhead{$\log P_B$} & \colhead{Anom} & \colhead{Var} &\colhead{ Exp} & \colhead{$E_{med}$}  \\

\colhead{\#} & \colhead{} &
\colhead{($\deg$)} & \colhead{($\deg$)} & \colhead{($\arcsec$)} & \colhead{($\arcmin$)} &
\colhead{(cts)} & \colhead{(cts)} & \colhead{(cts)} & \colhead{(cts)} & \colhead{frac} &
\colhead{} & \colhead{} & \colhead{} & \colhead{} & \colhead{(ks)} & \colhead{(keV)}
 \\

\colhead{(1)} & \colhead{(2)} &
\colhead{(3)} & \colhead{(4)} & \colhead{(5)} & \colhead{(6)} &
\colhead{(7)} & \colhead{(8)} & \colhead{(9)} & \colhead{(10)} & \colhead{(11)} &
\colhead{(12)} & \colhead{(13)} & \colhead{(14)} & \colhead{(15)} & \colhead{(16)} & \colhead{(17)}  }

\startdata
   2 & 063016.82$+$050452.2   & 97.570122 &   5.081179 &  1.4 & 10.0 &     10.6 &   4.7 &   6.4 &     1.1 & 0.90 &    2.0 & -3.4 & g... & \nodata &   11.8 & 1.4 \\
   3 & 063018.34$+$045953.9   & 97.576458 &   4.998324 &  1.0 &  8.0 &     9.8 &   4.3 &   4.2 &     7.8 & 0.90 &    2.0 & -3.9 & .... & a &   15.4 & 4.3 \\
   4 & 063020.40$+$050351.7   & 97.585038 &   5.064382 &  1.2 &  8.7 &      9.2 &   4.3 &   4.8 &     2.4 & 0.90 &    1.9 & -3.3 & .... & c &   14.8 & 1.7 \\
   5 & 063022.82$+$045538.2   & 97.595104 &   4.927297 &  1.0 &  7.8 &     8.2 &   4.0 &   3.8 &     7.3 & 0.89 &    1.8 & -3.2 & .... & a &   15.4 & 2.6 \\
   6 & 063023.75$+$050555.8   & 97.598962 &   5.098853 &  0.8 &  9.4 &     36.8 &   7.1 &   6.2 &    30.1 & 0.91 &    4.8 & $<$-5 & g... & \nodata &   10.5 & 3.4 \\
   8 & 063025.39$+$050540.5   & 97.605833 &   5.094589 &  1.0 &  8.9 &     15.2 &   5.0 &   4.8 &     2.2 & 0.90 &    2.7 & $<$-5 & g... & \nodata &   11.4 & 1.1 \\
  10 & 063026.78$+$045341.5   & 97.611598 &   4.894866 &  0.9 &  8.1 &      8.1 &   3.7 &   1.9 &     6.5 & 0.77 &    1.9 & -4.5 & g... & \nodata &   13.9 & 2.4 \\
  11 & 063027.33$+$045347.7   & 97.613897 &   4.896602 &  0.8 &  7.9 &     11.5 &   4.3 &   2.5 &     4.1 & 0.85 &    2.4 & $<$-5 & .... & a &   14.9 & 1.6 \\
  12 & 063028.13$+$050530.4   & 97.617210 &   5.091779 &  0.8 &  8.3 &     20.7 &   5.5 &   4.3 &     5.1 & 0.90 &    3.4 & $<$-5 & g... & \nodata &   14.1 & 1.7 \\
  14 & 063028.66$+$050137.8   & 97.619421 &   5.027193 &  0.3 &  5.9 &     32.6 &   6.3 &   1.4 &     2.1 & 0.90 &    4.7 & $<$-5 & .... & a &   16.6 & 1.2 \\
\enddata

\tablecomments{Table \ref{tbl:primary} is published in its entirety in the
electronic edition of the {\it Astrophysical Journal}.  A portion is
shown here for guidance regarding its form and content.}

\tablecomments{{\bf Column 1:} X-ray catalog sequence number, sorted by RA.
{\bf Column 2:} IAU designation.
{\bf Columns 3,4:} Right ascension and declination for epoch J2000.0.
{\bf Column 5:} Estimated random component of position error, $1\sigma$, in arcseconds computed as (standard deviation of PSF inside extraction region)/ $\sqrt{}$(\# of counts extracted).
{\bf Column 6:} Off-axis angle in arcminutes.
{\bf Columns 7,8:} Estimated net counts extracted in total energy band (0.5--8~keV) with Poisson-based standard errors.
{\bf Column 9:} Background counts extracted (total band).
{\bf Column 10:} Estimated net counts extracted in the hard energy band (2--8~keV).
{\bf Column 11:} Fraction of the PSF (at 1.497 keV) enclosed within the extraction region.  Note that a reduced PSF fraction (significantly below 90\%) may indicate that the source is in a crowded region.
{\bf Column 12:} Photometric significance computed as (net counts)/(upper error on net counts).
{\bf Column 13:} Log probability that extracted counts (total band) are solely from background.  Some sources have $P_B$ values above the 1\% threshold that defines the catalog because local background estimates can rise during the final extraction iteration after sources are removed from the catalog.
{\bf Column 14:}  Source anomalies:  g = fractional time that source was on a detector (FRACEXPO from {\em mkarf}) is $<0.9$ ; e = source on field edge; p = source piled up; s = source on readout streak.
{\bf Column 15:} Variability characterization based on the Kolmogorov-Smirnov statistic in the total band:  a = no evidence for variability ($0.05<P_{KS}$); b = possibly variable ($0.005<P_{KS}<0.05$); c = definitely variable ($P_{KS}<0.005$).  No value is reported for sources with fewer than 4 counts or for sources in chip gaps or on field edges.
{\bf Column 16:} Effective exposure time i ks: approximate time the source would have to be observed on axis to obtain the reported number of counts.
{\bf Column 17:} Background-corrected median photon energy (total band).}

\end{deluxetable}
\begin{deluxetable}{rcrrrrrrrrrrrccrr}
\centering \rotate \tabletypesize{\scriptsize} \tablewidth{0pt}
\tablecolumns{17}

\tablecaption{Tentative {\em Chandra} Catalog:  X-ray Source Properties \label{tbl:tentative}}

\tablehead{
\multicolumn{2}{c}{Source} &
\multicolumn{4}{c}{Position} &
\multicolumn{5}{c}{Extraction} &
\multicolumn{6}{c}{Characteristics} \\

\multicolumn{2}{c}{\hrulefill} &
\multicolumn{4}{c}{\hrulefill} &
\multicolumn{5}{c}{\hrulefill} &
\multicolumn{6}{c}{\hrulefill} \\

\colhead{Seq.} & \colhead{CXOU J} &
\colhead{$\alpha$} & \colhead{$\delta$} & \colhead{$\sigma$} & \colhead{$\theta$} &
\colhead{$C_{t}$} & \colhead{$\sigma_{t}$} & \colhead{$B_{t}$} & \colhead{$C_{h}$} & \colhead{PSF} &
\colhead{Signif} & \colhead{$\log P_B$} & \colhead{Anom} & \colhead{Var} &\colhead{ Exp} & \colhead{$E_{med}$}  \\

\colhead{\#} & \colhead{} &
\colhead{($\deg$)} & \colhead{($\deg$)} & \colhead{($\arcsec$)} & \colhead{($\arcmin$)} &
\colhead{(cts)} & \colhead{(cts)} & \colhead{(cts)} & \colhead{(cts)} & \colhead{frac} &
\colhead{} & \colhead{} & \colhead{} & \colhead{} & \colhead{(ks)} & \colhead{(keV)}
 \\

\colhead{(1)} & \colhead{(2)} &
\colhead{(3)} & \colhead{(4)} & \colhead{(5)} & \colhead{(6)} &
\colhead{(7)} & \colhead{(8)} & \colhead{(9)} & \colhead{(10)} & \colhead{(11)} &
\colhead{(12)} & \colhead{(13)} & \colhead{(14)} & \colhead{(15)} & \colhead{(16)} & \colhead{(17)}  }

\startdata
   1 & 063007.57$+$050440.7  &  97.531553 &   5.077982 &  1.9 & 11.9 &     9.0 &   5.5 &  14.0 &     0.9 & 0.90 &    1.5 & -1.8 & .... & a &   10.2 & 1.2 \\
   7 & 063025.18$+$045941.0   & 97.604953 &   4.994746 &  0.9 &  6.3 &      4.5 &   3.0 &   1.5 &     0.0 & 0.90 &    1.2 & -2.3 & .... & a &   16.7 & 1.1 \\
   9 & 063026.53$+$050240.9   & 97.610561 &   5.044707 &  1.1 &  6.8 &      4.0 &   3.0 &   2.0 &     0.5 & 0.89 &    1.1 & -1.7 & .... & a &   16.2 & 1.3 \\
  13 & 063028.23$+$045447.3   & 97.617645 &   4.913166 &  1.0 &  7.1 &      5.7 &   3.4 &   2.3 &     2.2 & 0.89 &    1.4 & -2.6 & g... & \nodata &   14.5 & 1.3 \\
  15 & 063028.91$+$045516.2   & 97.620469 &   4.921175 &  1.0 &  6.7 &      4.8 &   3.2 &   2.2 &     2.3 & 0.89 &    1.3 & -2.1 & .... & a &   16.6 & 2.1 \\
  22 & 063030.63$+$050223.4   & 97.627642 &   5.039857 &  0.8 &  5.8 &      4.4 &   3.0 &   1.6 &     3.8 & 0.90 &    1.2 & -2.2 & .... & a &   16.9 & 5.5 \\
  24 & 063032.45$+$050311.8   & 97.635212 &   5.053278 &  0.9 &  5.9 &      4.5 &   3.0 &   1.5 &     2.9 & 0.91 &    1.2 & -2.3 & g... & \nodata &   15.4 & 4.0 \\
  26 & 063032.75$+$045834.5   & 97.636484 &   4.976272 &  0.8 &  4.4 &      2.3 &   2.3 &   0.7 &     0.0 & 0.91 &    0.8 & -1.5 & g... & \nodata &   15.5 & 1.0 \\
  27 & 063032.83$+$050407.3   & 97.636818 &   5.068696 &  1.0 &  6.5 &      4.1 &   3.0 &   1.9 &     0.9 & 0.90 &    1.1 & -1.9 & .... & a &   15.4 & 1.2 \\
  29 & 063033.14$+$045704.6   & 97.638123 &   4.951299 &  0.8 &  4.8 &     3.3 &   2.5 &   0.7 &     2.5 & 0.89 &    1.0 & -2.2 & .... & a &   16.8 & 4.2 \\
\enddata

\tablecomments{Table \ref{tbl:tentative} is published in its entirety in the
electronic edition of the {\it Astrophysical Journal}.  A portion is
shown here for guidance regarding its form and content.}

\tablecomments{{\bf Column 1:} X-ray catalog sequence number, sorted by RA.
{\bf Column 2:} IAU designation.
{\bf Columns 3,4:} Right ascension and declination for epoch J2000.0.
{\bf Column 5:} Estimated random component of position error, $1\sigma$, in arcseconds computed as (standard deviation of PSF inside extraction region)/ $\sqrt{}$(\# of counts extracted).
{\bf Column 6:} Off-axis angle in arcminutes.
{\bf Columns 7,8:} Estimated net counts extracted in total energy band (0.5--8~keV) with Poisson-based standard errors.
{\bf Column 9:} Background counts extracted (total band).
{\bf Column 10:} Estimated net counts extracted in the hard energy band (2--8~keV).
{\bf Column 11:} Fraction of the PSF (at 1.497 keV) enclosed within the extraction region.  Note that a reduced PSF fraction (significantly below 90\%) may indicate that the source is in a crowded region.
{\bf Column 12:} Photometric significance computed as (net counts)/(upper error on net counts).
{\bf Column 13:} Log probability that extracted counts (total band) are solely from background.  Some sources have $P_B$ values above the 1\% threshold that defines the catalog because local background estimates can rise during the final extraction iteration after sources are removed from the catalog.
{\bf Column 14:}  Source anomalies:  g = fractional time that source was on a detector (FRACEXPO from {\em mkarf}) is $<0.9$ ; e = source on field edge; p = source piled up; s = source on readout streak.
{\bf Column 15:} Variability characterization based on the Kolmogorov-Smirnov statistic in the total band:  a = no evidence for variability ($0.05<P_{KS}$); b = possibly variable ($0.005<P_{KS}<0.05$); c = definitely variable ($P_{KS}<0.005$).  No value is reported for sources with fewer than 4 counts or for sources in chip gaps or on field edges.
{\bf Column 16:} Effective exposure time i ks: approximate time the source would have to be observed on axis to obtain the reported number of counts.
{\bf Column 17:} Background-corrected median photon energy (total band).}

\end{deluxetable}
\begin{deluxetable}{rcrrlllrrrrr}
\centering \rotate \tabletypesize{\scriptsize} \tablewidth{0pt}
\tablecolumns{12}

\tablecaption{X-ray Spectroscopy \label{tbl:thermal_spectroscopy}}

\tablehead{
\multicolumn{4}{c}{Source\tablenotemark{a}} &
\multicolumn{3}{c}{Spectral Fit\tablenotemark{b}} &
\multicolumn{5}{c}{X-ray Luminosities\tablenotemark{c}} \\

\multicolumn{4}{c}{\hrulefill} &
\multicolumn{3}{c}{\hrulefill} &
\multicolumn{5}{c}{\hrulefill} \\

\colhead{Seq. No.} & \colhead{CXOU J} & \colhead{$C_{t,net}$} & \colhead{Signif} &
\colhead{$\log N_H$} & \colhead{$kT$} & \colhead{$\log EM$} &
\colhead{$\log L_s$} & \colhead{$\log L_h$} & \colhead{$\log L_{h,c}$} & \colhead{$\log L_t$} & \colhead{$\log L_{t,c}$}  \\

\colhead{} & \colhead{} & \colhead{} & \colhead{} &
\colhead{(cm$^{-2}$)} & \colhead{(keV)} & \colhead{(cm$^{-3}$)} &
\multicolumn{5}{c}{(ergs s$^{-1}$)} \\

\colhead{(1)} & \colhead{(2)} & \colhead{(3)} & \colhead{(4)} &
\colhead{(5)} & \colhead{(6)} & \colhead{(7)} &
\colhead{(8)} & \colhead{(9)} & \colhead{(10)} &\colhead{(11)} & \colhead{(12)} 
}

\startdata
   2 & 063016.82$+$050452.2 &    10.6 &   2.0 &$ 22.0~({-0.3},{+0.2})$ & $\phantom{0.0} 0.4$ & $ 54.4({\cdots},{+0.3})$ &30.0 & 29.1 & 29.2 & 30.0 & 31.3 \\
   6 & 063023.75$+$050555.8 &    36.8 &   4.8 &$ 22.4$ & $\phantom{0.0} 12.8$ & $ 54.2~({-0.2},{+0.2})$ &29.82 & 31.2 & 31.2 & 31.2 & 31.4 \\
   8 & 063025.39$+$050540.5 &    15.2 &   2.7 &$ 21.3~({\cdots},{+0.4})$ & $\phantom{0.0} 1.0~({-0.2},{+0.2})$ & $ 53.5$ &30.2 & 29.5 & 29.5 & 30.3 & 30.5 \\
  11 & 063027.33$+$045347.7 &    11.5 &   2.4 &$ 21.9$ & $\phantom{0.0} 1.2$ & $ 53.6~({\cdots},{+0.2})$ &29.84 & 29.7 & 29.8 & 30.1 & 30.6 \\
  12 & 063028.13$+$050530.4 &    20.7 &   3.4 &$ 21.2~({\cdots},{+0.6})$ & $\phantom{0.0} 2.1$ & $ 53.5~({-0.3},{+0.3})$ &30.1 & 30.1 & 30.1 & 30.4 & 30.5  \\
  14 & 063028.66$+$050137.8 &    32.6 &   4.7 &$ 21.9$ & $\phantom{0.0} 0.6~({-0.3},{+0.7})$ & $ 54.3~({\cdots},{+1.0})$ &30.3 & 29.6 & 29.7 & 30.4 & 31.3  \\
  16 & 063028.95$+$050512.4 &    10.3 &   2.1 &$ 20.4$ & $\phantom{0.0} 5.8$ & $ 53.2$ &29.9 & 30.1 & 30.1 & 30.3 & 30.3  \\
  17 & 063029.28$+$050113.2 &     9.8 &   2.2 &$ 22.1$ & $\phantom{0.0} 10.7$ & $ 53.3$ &29.3 & 30.3 & 30.4 & 30.3 & 30.5  \\
  19 & 063029.51$+$050053.1 &    11.7 &   2.5 &$ 22.1~({\cdots},{+0.5})$ & $\phantom{0.0} 1.1$ & $ -5.8$ & 29.29 &  30.5 &  30.5 &  30.5 &  30.6 \\
 20 & 063030.14$+$045949.3 &    14.1 &   2.8 &$ 23.0~({-0.3},{+0.4})$ & $\phantom{0.0} 3.4$ & $ 54.3~({\cdots},{+1.3})$ &  \nodata & 30.7 & 31.1 & 30.7 &   \nodata \\
\enddata

\tablenotetext{a}{
Cols.\ (1)--(4) reproduce the source identification, net counts, and photometric significance data from Table~\ref{tbl:primary}.
}

\tablenotetext{b}{
Cols.\ (5) and (6) present the best-fit values for the extinction column density and plasma temperature parameters.  In four cases (\#19, 113, 162 and 163), a powerlaw spectral model
gives a significantly better fit than a single-temperature thermal model.  In these cases,
column (6) gives the powerlaw index $\Gamma$ and column (7) gives the normalization
of the spectrum in italics.
Col.\ (7) presents the emission measure derived from the model spectrum, assuming a distance of 1.4~kpc.
Quantities marked with an asterisk (*) were frozen in the fit.
Uncertainties represent 90\% confidence intervals.
More significant digits are used for uncertainties $<$0.1 in order to avoid large rounding errors; for consistency, the same number of significant digits is used for both lower and upper uncertainties.
Uncertainties are missing when XSPEC was unable to compute them or when their values were so large that the parameter is effectively unconstrained.
Fits lacking uncertainties, fits with large uncertainties, and fits with frozen parameters should be viewed merely as splines to the data to obtain rough estimates of luminosities; the listed parameter values are unreliable.
}

\tablenotetext{c}{ X-ray luminosities derived from the model spectrum are presented in cols.\ (8)--(12): (s) soft band (0.5--2 keV); (h) hard band (2--8 keV); (t) total band (0.5--8 keV).
Absorption-corrected luminosities are subscripted with a $c$.
Cols. (8) and (12) are omitted when $\log N_H > 22.5$~cm$^{-2}$ since the soft band emission is essentially unmeasurable.
}

\end{deluxetable}
\begin{deluxetable}{rcccrrrccrrrc}
\centering \rotate
\tabletypesize{\scriptsize} \tablewidth{0pt}
\tablecolumns{13}

\tablecaption{Stellar Counterparts \label{tbl:counterparts}}
\tablehead{
\multicolumn{2}{c}{X-ray Source} && \multicolumn{10}{c}{Optical/Infrared Photometry} \\
\cline{1-2} \cline{4-13}

\colhead{Seq} & \colhead{CXOU J} & \colhead{} & \colhead{USNO B1.0} & \colhead{B} & \colhead{R} & \colhead{I} & \colhead{2MASS} & \colhead{FLAMINGOS} & \colhead{J} & \colhead{H} & \colhead{K} & \colhead{PhCcFlg} \\

\colhead{\#} & \colhead{} & \colhead{} & \colhead{ID} & \colhead{(mag)} & \colhead{(mag)} & \colhead{(mag)} & \colhead{ID} & \colhead{ID} & \colhead{(mag)} & \colhead{(mag)} & \colhead{(mag)} & \colhead{}  \\

\colhead{(1)} & \colhead{(2)} & \colhead{} & \colhead{(3)} & \colhead{(4)} &
\colhead{(5)} & \colhead{(6)} & \colhead{(7)} &
\colhead{(8)} & \colhead{(9)} & \colhead{(10)} &\colhead{(11)} & \colhead{(12)}
}

\startdata
  1 & 063007.57+050440.7 & & \nodata & \nodata & \nodata & \nodata & 06300735+0504423 & 063007+050442 & 14.12 & 13.26 & 12.89 & AAA000 \\
  2 & 063016.82+050452.2 & & 0950-0096116 & 20.03 & 17.53 & 16.05 & 06301680+0504518 & 063016+050452 & 14.52 & 13.74 & 13.42 & AAA000 \\
  3 & 063018.34+045953.9 & & \nodata & \nodata & \nodata & \nodata & \nodata & \nodata & \nodata & \nodata & \nodata & \nodata\nodata \\
  4 & 063020.40+050351.7 & & 0950-0096157 & 20.41 &  \nodata & 16.90 & 06302046+0503528 & 063020+050352 & 15.04 & 14.25 & 13.89 & AAA000 \\
  5 & 063022.82+045538.2 & & \nodata & \nodata & \nodata & \nodata & \nodata & 063022+045539 & \nodata & 18.36 & 17.43 & \nodata\nodata \\
  6 & 063023.75+050555.8 & & \nodata & \nodata & \nodata & \nodata & \nodata & \nodata & \nodata & \nodata & \nodata & \nodata\nodata \\
  7 & 063025.18+045941.0 & & 0949-0093548 & 20.41 & 18.04 & 16.33 & 06302519+0459412 & 063025+045941 & 14.80 & 14.20 & 13.86 & AAA000 \\
  8 & 063025.39+050540.5 & & 0950-0096215 & 18.79 & 16.75 & 15.39 & 06302550+0505402 & 063025+050540 & 13.89 & 13.15 & 12.81 & AAA000 \\
  9 & 063026.53+050240.9 & & 0950-0096223 &  \nodata & 18.50 & 17.32 & 06302649+0502411 & 063026+050241 & 15.61 & 14.74 & 14.37 & AAA000 \\
 10 & 063026.78+045341.5 & & \nodata & \nodata & \nodata & \nodata & \nodata & 063026+045344 & \nodata & 18.22 & 17.45 & \nodata\nodata \\
\enddata

\tablecomments{Table \ref{tbl:counterparts} is published in its entirety in the
electronic edition of the {\it Astrophysical Journal}.  A portion is
shown here for guidance regarding its form and content.}

\tablecomments{The errors of the $BRI$ magnitudes are $<0.02$ mag in
  general for stars $V<17$ mag and become $>0.05$ mag for fainter
  stars $V>19$ mag (Bergh{\"o}fer \& Christian\ 2002).  The mean
  photometric uncertainties in FLAMINGOS are $0.058\pm 0.012$ ,
  $0.064\pm 0.018$ , and $0.056\pm 0.014$ for J, H, and K,
  respectively \citep{RomanZuniga08}. The typical errors are much
  smaller for brighter stars (e.g., $\sigma_J=0.02$ mag for stars with
  $J<16$ mag). For 2MASS photometric quality flags, see Cutri et
  al. (2003).}

\tablecomments{{\bf Columns 1--2} reproduce the sequence number and
source identification from Table~\ref{tbl:primary} and
Table~\ref{tbl:tentative}. For convenience, [OI81]=Ogura \& Ishida
(1981), [MJD95]=Massey, Johnson, \& Degioia-Eastwood (1995),
[BC02]=Bergh{\"o}fer \& Christian\ (2002)}

\tablenotetext{      53}{=[OI81] 14 =[MJD95] 104; spectral type B1V; pmRA=11.0 mas/yr, pmDE=-2.8 mas/yr}
\tablenotetext{      54}{=[OI81] 10 =[MJD95] 108; spectral type B2V; pmRA=-2.3 mas/yr, pmDE=-11.9 mas/yr}
\tablenotetext{      61}{=V539 Mon =[OI81] 13 =[MJD95] 110=MSX6C G206.1821-02.3456; pmRA=2.8 mas/yr, pmDE=0.4 mas/yr}
\tablenotetext{      71}{=[OI81] 12 =[MJD95] 102; pmRA=6.8 mas/yr, pmDE=0.6 mas/yr}
\tablenotetext{      128}{=[OI81] 35 =[MJD95] 471; spectral type A2:; pmRA=-0.8 mas/yr, pmDE=3.6 mas/yr}
\tablenotetext{      138}{=[OI81] 36 =[MJD95] 497; spectral type B5; pmRA=6.5 mas/yr, pmDE=2.1 mas/yr}
\tablenotetext{      141}{=[MJD95] 498; pmRA=-3.0 mas/yr, pmDE=1.9 mas/yr}
\tablenotetext{      149}{=[BC02] 11; known X-ray source; $\log Lx$(ROSAT/PSPC)=31.01 erg s$^{-1}$; pmRA=0.6 mas/yr, pmDE=-12.6 mas/yr}
\tablenotetext{      161}{=[MJD95] 653; pmRA=-1.0 mas/yr, pmDE=-5.4 mas/yr}
\end{deluxetable}

\clearpage


\newpage
\pagestyle{empty}
\includegraphics[width=9in,angle=90]{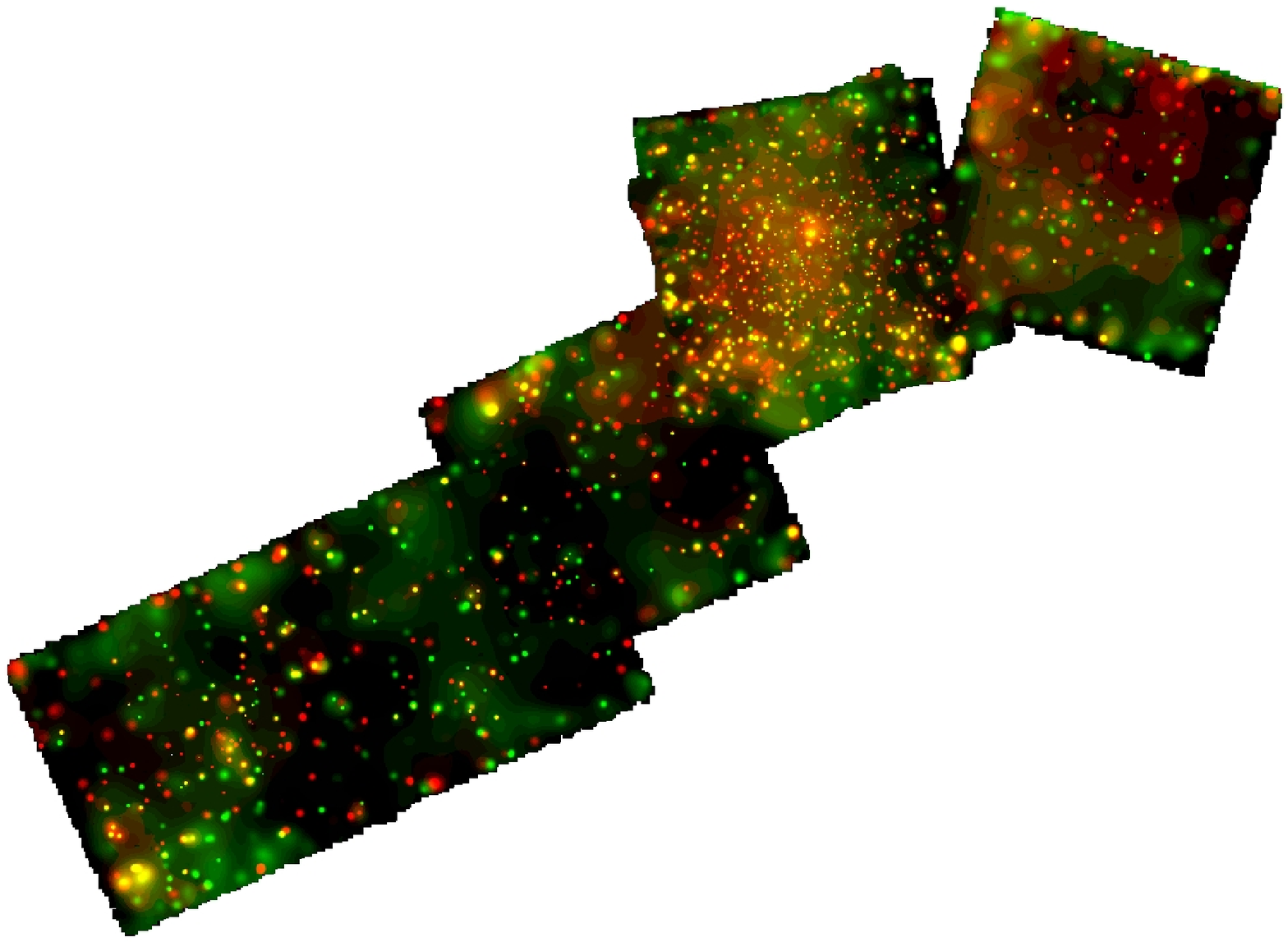}

\begin{figure}[H]
\plotone{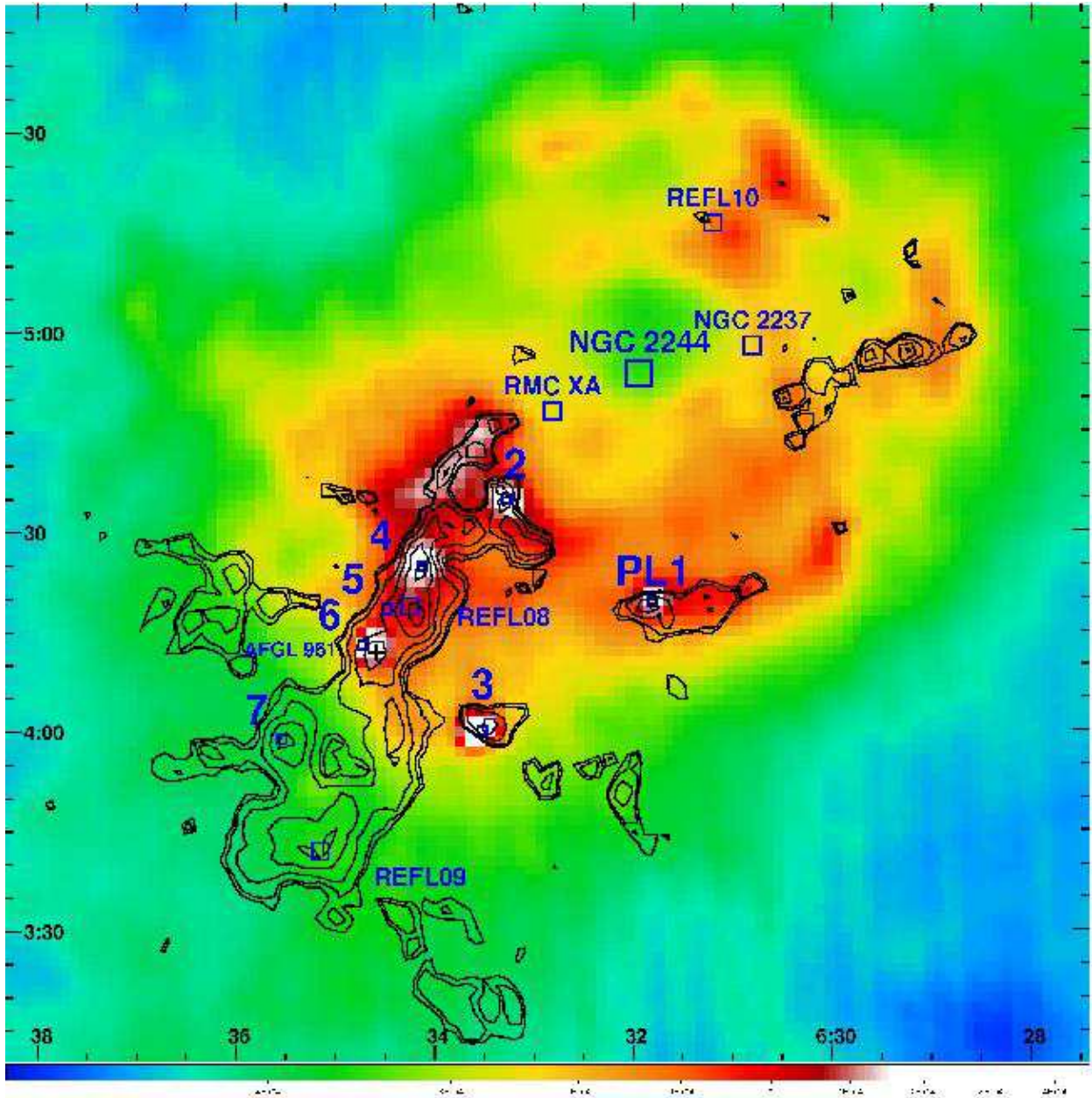}
\caption{(a) A large scale view of the Rosette Complex using the
  100$\mu$m IRAS map shown in color. $^{13}$CO emission contours from
  \citet{Heyer06} are overlaid.  Previously identified clusters from
  \citet{PL97}, \citet{RomanZuniga08}, and \citet{Wang08b} are
  labeled.  (b) A cartoon presentation of (a) with the dark clouds and
  nearby supernova remnant outlined and the clusters marked with
  stars.\label{fig:mosaic2}}
\end{figure}
\newpage
\centerline{\includegraphics[angle=-90,width=0.9\textwidth]{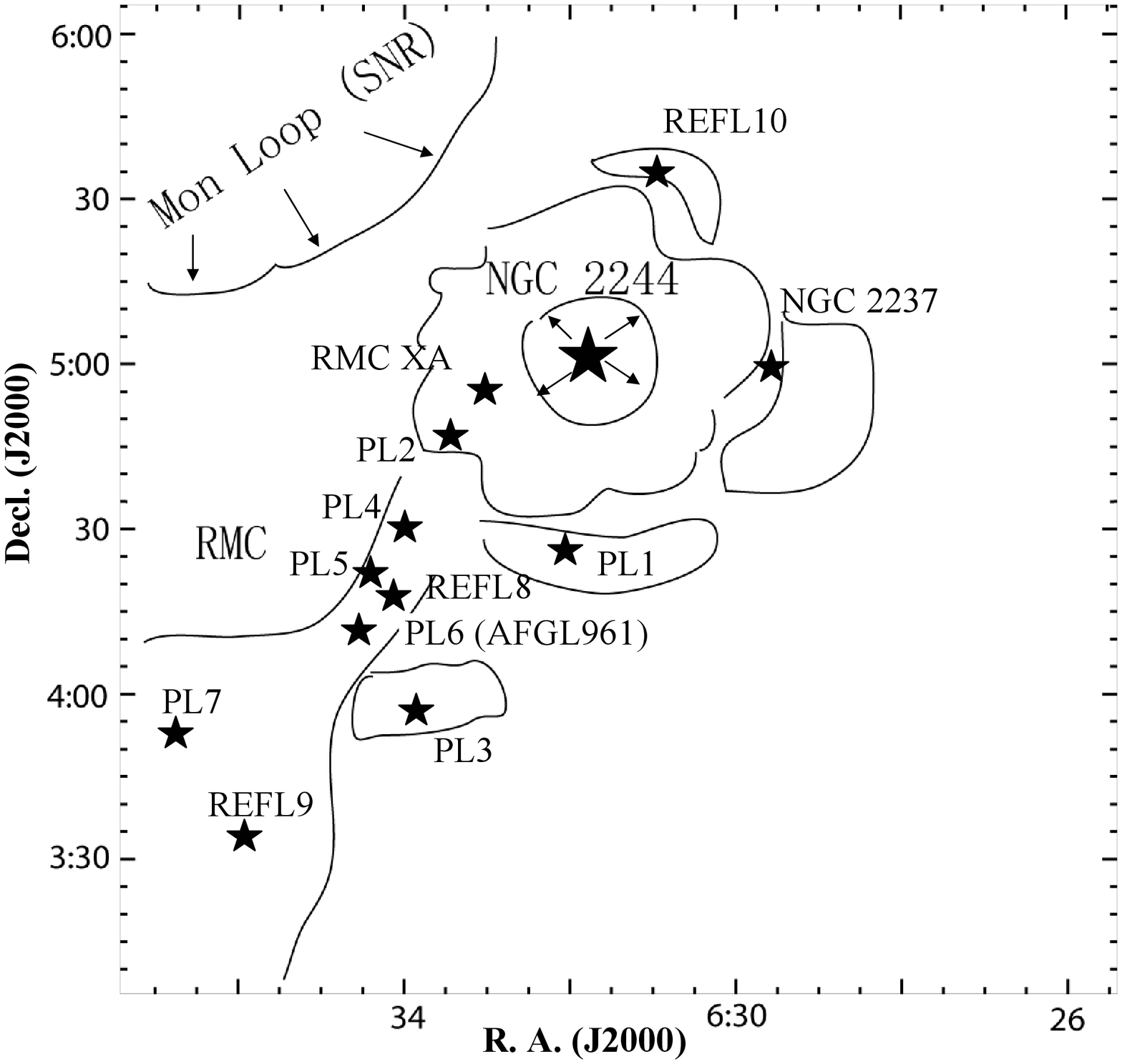}}
\centerline{Figure~\ref{fig:mosaic2}(b)---Continued.}

\end{document}